\documentclass[12pt,12pt]{article}
\usepackage{graphicx}
\usepackage{epsfig}
\usepackage{amsmath,amssymb}
\usepackage{slashed} 
\catcode`\@=11
\newdimen\ex@
\font\dozeb=cmmib10 scaled \magstep1
\font\dozesyb=cmbsy10 scaled \magstep1
\font\dezb=cmmib10
\def\bm{\fam9}
\font\mathbf cmbxti10 at 12pt

\def\beq{\begin{equation}}
\def\eeq{\end{equation}}
\def\beqa{\begin{eqnarray}}
\def\eeqa{\end{eqnarray}}
\newcommand{\ba}{\begin{eqnarray}}
\newcommand{\ea}{\end{eqnarray}}
\newcommand\BA{\begin{array}}
\newcommand\EA{\end{array}}

\def\D{{\bm D}}
\def\E{{\bm E}}
\def\H{{\bm H}}
\def\K{{\bm K}}

\begin{document}
\def\thefootnote{\fnsymbol{footnote}}

\title{\bf Modified Non-Euclidian Transformation \\ 
on the
$\frac{\mbox{\bf SO(2N+2)}}{\mbox{\bf U(N+1)}}$ Grassmannian and \\
{\bf SO(2N+1)} Random Phase Approximation\\ for Unified Description of \\
Bose and Fermi Type Collective Excitations\footnotemark[1]}

\author
{Seiya NISHIYAMA\footnotemark[2] $\!\!$, 
Jo\~ao da PROVID\^{E}NCIA\footnotemark[3]\\
\\
Centro de F\'\i sica,
Departamento de F\'\i sica,\\ 
Universidade de Coimbra,
P-3004-516 Coimbra, Portugal\\
\\[-0.01cm]
{\it Dedicated to the Memory of Hideo Fukutome}}

\maketitle

\footnotetext[1]
{A part of this work is a contributed paper presented by H. Fukutome and S. Nishiyama
at the March 1982 Sanibel Symposium on
{\it Atomic, Molecular and Solid- State Theory, Collision Phenomena
and Computational Quantum Chemistry},
Flagler Beach, Florida.

~~~Particularly, sections 5 and 6 owe to private communications
with the late Professor H.  Fukutome.}
\footnotetext[2]
{Corresponding author.
~E-mail address: seikoceu@khe.biglobe.ne.jp,~
nisiyama@teor.fis.uc.pt}
\footnotetext[3]
{E-mail address: providencia@teor.fis.uc.pt}

\vspace{-9mm}   

\begin{abstract}
$\!\!\!\!$In a slight different way from the previous one,
we propose a modified non-Euclidian transformation
on the $\frac{SO(2N \!+\! 2)}{U(N \!+\! 1)}$ Grassmannian
which give the projected $SO(2N \!+\! 1)$ Tamm-Dancoff equation.
We derive a classical time dependent (TD) $\!SO(2N \!+\! 1)\!$ Lagrangian
which,
through the Euler-Lagrange equation
of motion for $\!\frac{SO(2N \!\!+\!\! 2)}{U(N \!\!+\!\! 1)}\!$ coset variables,
brings
another form of the previous extended-TD $\!$Hartree-Bogoliubov (HB) equation.
The $SO(2N \!\!+\!\! 1)$ random phase approximation (RPA) is derived
using Dyson representation for paired and unpaired operators.
In the $\!SO(2N)\!$ HB case,
one boson and two boson excited states are realized.
We, however,  stress non existence of a higher RPA vacuum.
An integrable system is given by a geometrical concept of zero-curvature, i.e.,
integrability condition of connection on
the corresponding Lie group.
From the group theoretical viewpoint,
we show the existence of a symplectic two-form
$\omega$.
\end{abstract}
\vspace{-0.1cm}
~~~~~~~Keywords: Hartree-Bogoliubov formalism; 
SO(2N) and SO(2N+1) Lie algebras; 

~~~~~~~~~TD Hartree-Bogoliubov equation;
SO(2N+1) random phase approximation \\
\vspace{0.1cm} 
 ~~~~~~Mathematics Subject Classification 2010: 81Vxx, 81V35, 81V70


\newpage

\setcounter{equation}{0}
\renewcommand{\theequation}{\arabic{section}.\arabic{equation}}

\section{Introduction}

\vspace{-0.3cm}

~~~~
The time dependent Hartree-Bogoliubov (TDHB) theory is the first standard
approximation in the many-body theoretical description
of a superconducting fermion system
\cite{Bogo.59,BCS.57}.
It is a good approximation for
the ground state of a fermion system with a pairing interaction.
The $SO(2N)$ Lie algebra of the fermion pair operators contains
the $U(N)$ Lie algebra as a subalgebra. 
$SO(2N)$ and $U(N)$ denote the special orthogonal group 
of $2N$ dimensions and the unitary group of $N$ dimensions
($N$: Number of single-particle states of the fermions). 
The canonical transformation
of the fermion operators generated by the Lie operators 
in the $SO(2N)$ Lie algebra induces
the generalized Bogoliubov transformation for the fermions.
The  Euler-Lagrange equation of motion
for $\!\!\frac{SO\!(2N)}{U\!(N)}\!$ coset variables
makes a TDHB equation$\!\!$
\cite{Nishi.81}.

For providing a general microscopic
means for a unified self-consistent (SC) description for Bose and Fermi type
collective excitations in such fermion systems, 
a new many-body theory has been
proposed by Fukutome, Yamamura and one of the present authors (S.N.)
standing on the $SO(2N\!+\!1)$ Lie algebra of the fermion operators
\cite{Fuku.Yama.Nishi.77}. 
An induced representation of an $SO(2N\!+\!1)$ group has been obtained
from a group extension of the $SO(2N)$ Bogoliubov transformation for fermions
to a new canonical transformation group.
We start with the fact that the set of the fermion operators consisting of
creation-annihilation and pair operators forms a larger Lie algebra,
the Lie algebra of the $SO(2N\!+\!1)$ group.
The fermion Lie operators, when operated onto the
integral representation of the $SO(2N\!+\!1)$ wave function  (HW), are mapped into
the regular representation of the $SO(2N\!+\!1)$ group and are represented
by Bose operators. 
The creation-annihilation operators themselves as well as the pair 
operators are given by the Schwinger type boson representation
\cite{Schwin.65,Yama.Nishi.76}.
Embedding an $SO(2N\!\!+\!\!1)$ group into an $SO(2N\!+\!2)$ group
and using
$\frac{SO(2N\!+\!2)}{U(N\!+\!1)}$ coset variables
\cite{Fuku.77,Doba1.81,Doba2.81,Fuku.Nishi.84},
we have developed an extended TDHB theory. 
This extended TDHB theory applicable to both even and odd fermion-number systems
is a SC field (SCF) theory with the same level
of the mean field approximation
as the usual TDHB theory for even fermion-number systems.

$\!\!$With a slight different way from the Fukutome's
\cite{Fuku.78},
we propose a modified non-Euclidian transformation
on the $\frac{SO(2N \!+\! 2)}{U(N \!+\! 1)}$ Grassmannian
which give the projected $\!SO(2\!N \!+\!1)\!$ Tamm-Dancoff equation$\!$
\cite{Fuku.78}.
We derive a classical TD $SO(2N \!\!+\!\! 1)$ Lagrangian
which,
by through the Euler-Lagrange equation
of motion for $\!\frac{SO(2N \!\!+\!\! 2)}{U(N \!\!+\!\! 1)}\!$ coset variables,
brings
another form of the previous extended-TDHB equation.
The $SO(2N \!\!+\!\! 1)$ random phase approximation (RPA) is derived
using Dyson representation (rep)
\cite{Fuku.77}
for paired and unpaired operators.
In the $SO(2N)$ HB case,
a RPA vacuum is obtained and
one boson and two boson excited states are realized.
We, however, stress non existence of a higher RPA vacuum
because an $SO(2N \!\!+\!\! 1)$
spinor function is not integrable.
An integrable system is given by a geometrical concept of zero-curvature, i.e.,
integrability condition of connection on
the corresponding Lie group.
From the group theoretical viewpoint,
we show the existence of a symplectic two-form
$\omega$.

$\!\!$This paper is organized as follows:
First we give a summary of embedding of $SO(2N \!\!+\!\! 1)$ Bogoliubov transformation
into $\!SO(2N \!\!+\!\! 2)\!$ group.
In section 3
we provide a differential form for boson 
over $\frac{SO(2N\!+\!2)}{U(N\!+\!1)}$ coset space
which brings Dyson rep for paired and unpaired operators.
In section 4
we give a matrix-valued generator coordinate and non-Euclidian transformation.
In section 5
we derive the classical TD $SO(2N \!\!+\!\! 1)$ Lagrangian.
In section 6
the RPA vacuum is obtained
using the Dyson rep
and the existence of the symplectic two-form
$\omega$
is shown.
Finally, in the last section, we give some discussions and further perspective.


\newpage

\setcounter{equation}{0}
\renewcommand{\theequation}{\arabic{section}.\arabic{equation}}

\section{Summary of embedding of SO(2N $\!\!+\!\!$ 1) Bogoliubov transformation
 into SO(2N $\!\!+\!\!$ 2) group}

\def\bra#1{{<\!#1\,|}} 
\def\ket#1{{|\,#1\!>}}

~~~~Let $c_{\alpha }$ and $c^{\dag }_{\alpha }$, $\alpha$ 
\!=\! 
1,$\cdot \cdot \cdot$, 
$N$, be annihilation and creation operators of the fermion
satisfying the canonical anti-commutation relations\\[-12pt]
\beq
\{c_{\alpha },~c^{\dag }_{\beta }\}
=
\delta_{\alpha \beta } ,~~
\{c_{\alpha },~c_{\beta }\}
=
\{c^{\dag }_{\alpha },~c^{\dag }_{\beta }\}
=0 .
\label{anticommrel}
\eeq
We introduce the following
annihilation and creation operators and pair operators:\\[-18pt] 
\beqa
\left.
\BA{ll}
&c_{\alpha },~c^{\dag }_{\alpha } ,~~
E^{\alpha }_{~\beta }
=
c^{\dag }_{\alpha }c_{\beta }
{\displaystyle -\frac{1}{2}} \delta_{\alpha \beta } ,~~
E^{\alpha \beta }
=
c^{\dag }_{\alpha }c^{\dag }_{\beta } ,~~
E_{\alpha \beta }
=
c_{\alpha }c_{\beta } ,\\
\\[-10pt] 
&E^{\alpha \dag }_{~\beta }
=
E^{\beta }_{~\alpha } ,~~
E^{\alpha \beta } 
=
E^{\dag }_{\beta \alpha } ,~~
E_{\alpha \beta }
=
- E_{\beta \alpha } .~~
(\alpha , \beta = 1, \cdot \cdot \cdot, N)
\EA
\right\}
\label{operatorset}
\eeqa\\[-10pt]
The fermion operators
(\ref{operatorset})
form an $SO(2N \!+\! 1)$ Lie algebra.
Due to the anti-commutation relation
(\ref{anticommrel}),
the commutation relations for the operators
in the $SO(2N \!+\! 1)$ Lie algebra are\\[-20pt] 
\beqa
[E^{\alpha }_{~\beta },~E^{\gamma }_{~\delta }]
=
\delta_{\gamma \beta }E^{\alpha }_{~\delta } 
- 
\delta_{\alpha \delta }E^{\gamma }_{~\beta },~~
(U(N)~\mbox{algebra})
\label{commurel1}
\eeqa
\vspace{-1.0cm}
\beqa
\left.
\BA{ll}
&[E^{\alpha }_{~\beta },~E_{\gamma \delta }]
=
\delta_{\alpha \delta }E_{\beta \gamma } 
- 
\delta_{\alpha \gamma }E_{\beta \delta },~~
[E_{\alpha \beta },~E_{\gamma \delta }]
=
0 ,\\
\\[-10pt] 
&[E^{\alpha \beta },~E_{\gamma \delta }]
=
\delta_{\alpha \delta }E^{\beta }_{~\gamma } 
+ 
\delta_{\beta \gamma }E^{\alpha }_{~\delta }
-
\delta_{\alpha \gamma }E^{\beta }_{~\delta } 
- 
\delta_{\beta \delta }E^{\alpha }_{~\gamma }
\EA
\right\}
\label{commurel2}
\eeqa
\vspace{-0.6cm}
\beqa
\left.
\BA{ll}
&[c^{\dag }_{\alpha },~c_{\beta }]
=
2E^{\alpha }_{~\beta },~
[c_{\alpha },~c_{\beta }]
=
2E_{\alpha \beta },\\
\\[-10pt] 
&[c_{\alpha },~E^{\beta }_{~\gamma }]
=
\delta_{\alpha \beta }c_{\gamma },~
[c_{\alpha },~E_{\beta \gamma }]
=
0,\\
\\[-10pt] 
&[c_{\alpha },~E^{\beta \gamma }]
=
\delta_{\alpha \beta }c^{\dag }_{\gamma }
-
\delta_{\alpha \gamma }c^{\dag }_{\beta } .
\EA
\right\}
\label{commurel3}
\eeqa\\[-12pt]
We omit the commutation relations obtained 
by hermitian conjugation of
(\ref{commurel2}) and (\ref{commurel3}).
The $SO(2N \!+\! 1)$ Lie algebra of the fermion operators contains
the $U(N)(=\{E^{\alpha }_{~\beta }\})$ and 
the $SO(2N)(=\{E^{\alpha }_{~\beta },
~E^{\alpha \beta },~E_{\alpha \beta }\})$ 
Lie algebras of the pair operators 
as sub-algebras.

An $SO(2N)$ canonical transformation $U(g)$
is the generalized Bogoliubov transformation 
\cite{Bogo.59} 
specified by an $SO(2N)$ matrix $g$\\[-12pt]
\beq
U(g)(c, c^{\dag })U^{\dag }(g)
=
(c, c^{\dag }) g ,~~
g
\stackrel{\mathrm{def}}{=}
\left[ 
\BA{cc} 
a & \overline{b} \\
b &  \overline{a} \\ 
\EA 
\right] ,
~~
g^{\dag }g = gg^{\dag } = 1_{2N} ,~
\det g
=
1 ,
\label{Bogotrans}
\eeq
\beq
U(g)U(g') = U(gg') ,~~~
U(g^{-1}) = U^{-1}(g) = U^{\dag }(g) ,~~~
U(1_{2N}) = \mathbb{I}_g~(\mbox{unit operator on}~g).
\label{Ug}
\eeq
($c$, $c^{\dag }$) is 
a 2$N$-dimensional row vector 
(($c_{\alpha }$), ($c^{\dag }_{\alpha }$)). 
$a \!=\! (a^{\alpha }_{~\beta })$ and $b \!=\! (b_{\alpha \beta })$ 
are $N \!\times\! N$ matrices. 
The bar denotes the complex conjugation.
The HB ($SO(2N)$) WF $\ket g$ is generated as 
$\ket g \!=\! U(g) \ket 0$ 
($\ket 0$: vacuum satisfying 
$c_{\alpha }\ket 0 \!=\! 0$). 
The matrix $g$ is composed of matrices the $a$ and $b$ satisfying 
the ortho-normalization condition.
The $\! \ket g \!$ is expressed as follows:\\[-22pt]
\beqa 
\ket g
=
\bra 0 U(g) \ket 0
\exp(\frac{1}{2}\cdot q_{\alpha \beta }c^\dagger_\alpha 
c^\dagger_\beta) \ket 0 ,
\label{Bogoketg}
\eeqa
\vspace{-0.9cm}
\beqa
\bra 0 U(g) \ket 0
=
\overline{\Phi }_{00}(g)
=
\left[\det(a)\right]^{\frac{1}{2}}
=
\left[\det(1_N + q^\dag q)\right]^{-\frac{1}{4}}
e^{i\frac{\tau }{2}}~, 
\label{Bogowf}
\eeqa
\vspace{-1.0cm}
\beqa
q = ba^{-1} = -q^{\mbox{\scriptsize T}},~
\mbox{(variables of the $\frac{SO(2N)}{U(N)}$ coset space)} ,~
\tau 
=
\frac{i}{2} \ln \!
\left[\frac{\det(\overline{a})}{\det({a})}
\right] ,
\label{Bogocoset}
\eeqa\\[-18pt]
where $\det$ means determinant and the 
symbol {\scriptsize T} denotes the transposition. 

The canonical anti-commutation relation gives us not only 
the above Lie algebras but also the other three algebras.
Let $n$ be the fermion number operator 
$n \!=\! c^\dag _{\alpha } c_\alpha$.
The operator $(-1)^n$ anticommutes with 
$c_\alpha$ and $c^\dag _\alpha$;\\[-14pt]
\beq
\{ c_\alpha,~(-1)^n \}
=
\{ c^\dag _\alpha,~(-1)^n \}
=
0.
\label{chiralop}
\eeq

$\!\!\!\!\!\!$Introduce an operator
$
\Theta
$
defined by
$
\Theta
\!\!\equiv\!\!
\theta_\alpha c^\dag_\alpha \!-\! \overline{\theta }_\alpha c_\alpha 
$.
Due to the relation
$
\Theta ^2
\!\!=\!\!
-
\overline{\theta }_\alpha \theta_\alpha
$,
we have
\beqa\\[-16pt]
\left.
\BA{ll}
e^\Theta
=
Z 
+ X_\alpha c^\dag_\alpha
- \overline{X}_\alpha c_\alpha ,~~
\overline{X}_\alpha X_\alpha + Z^2 = 1 ,\\
\\[-12pt]
Z
=
\cos \theta ,~~
X_\alpha
=
{\displaystyle \frac{\theta_\alpha }{\theta }} \sin \theta ,~~
\theta ^2
=
\overline{\theta }_\alpha \theta_\alpha .
\EA
\right\}
\label{theta}
\eeqa\\[-14pt]
From
(\ref{anticommrel}), (\ref{chiralop}) and (\ref{theta}),
we obtain\\[-20pt]
\beqa
\left.
\BA{ll}
&e^\Theta (c_\alpha, c^\dag _\alpha ,
{\displaystyle \frac{1}{\sqrt{2}}}) (-1)^n e^{-\Theta }
=
(c_\beta, c^\dag _\beta ,
{\displaystyle \frac{1}{\sqrt{2}}}) (-1)^n G_X ,\\
\\[-6pt]
&G_X 
\stackrel{\mathrm{def}}{=}
\left[ 
\BA{ccc} 
\delta_{\beta \alpha } 
- 
\overline{X}_\beta X_\alpha &
\overline{X}_\beta \overline{X}_\alpha & -\sqrt{2}Z \overline{X}_\beta \\
\\[-8pt]
X_\beta X_\alpha & \delta_{\beta \alpha } 
- 
X_\beta \overline{X}_\alpha & 
\sqrt{2}ZX_\beta  \\
\\[-8pt]
\sqrt{2}ZX_\alpha & -\sqrt{2}Z \overline{X}_\alpha & 2Z^2 - 1 
\EA 
\right] .
\EA
\right\}
\label{chiraloptrans}
\eeqa\\[-12pt]
Let $G$ be the $(2N+1) \times (2N+1)$ matrix defined by\\[-16pt]
\beqa
\!\!\!\!\!\!\!\!
\left.
\BA{ll}
&G
\stackrel{\mathrm{def}}{=}
G_X \!
\left[ \!
\BA{ccc} 
a & \overline{b} & 0 \\
\\[-8pt]
b & \overline{a} & 0 \\
\\[-8pt]
0 & 0 & 1
\EA \!
\right]
\!=\!  
\left[ \!
\BA{ccc} 
a - \overline{X} Y & \overline{b} + \overline{X} \overline{Y} & -\sqrt{2}Z \overline{X} \\
\\[-8pt]
b + X Y & \overline{a} - X \overline{Y} & \sqrt{2}ZX  \\
\\[-8pt]
\sqrt{2}ZY & -\sqrt{2}Z \overline{Y} & 2Z^2 - 1
\EA \!
\right] ,
\BA{c}
X_\alpha
\!=\!
\overline{a}^{\alpha }_{~\beta } Y_\beta 
- 
b_{\alpha \beta } \overline{Y}_\beta ,\\
\\[-8pt]
Y_\alpha
\!=\!
X_\beta a^\beta_{~\alpha } 
- 
\overline{X}_\beta b_{\beta \alpha } ,\\
\\[-8pt]
\overline{Y}_\alpha Y_\alpha + Z^2 = 1 ,
\EA
\EA \!\!
\right\}
\label{defG} 
\eeqa\\[-12pt]
where $X$ and $Y$ are the column vector and the row vector, respectively.
The $SO(2N \!\!+\!\! 1)$ canonical transformation $U(G)$ is generated by 
the fermion $SO(2N \!\!+\!\! 1)$ Lie operators.
The $U(G)$ is an extension of 
the generalized Bogoliubov transformation $U(g)$ 
\cite{Bogo.59} 
to a nonlinear transformation and is specified 
by the $SO(2N \!\!+\!\! 1)$ matrix $G$.
We identify this $G$ with the argument $G$ of $U(G)$.
Then $U(G) \!=\! U(G_X) U(g)$ and $U(G_X) \!=\! \exp (\Theta )$.

From
(\ref{Bogotrans}), (\ref{chiraloptrans}) and (\ref{defG})
and the commutability of $U(g)$ with $(-1)^n$,
we obtain\\[-16pt]
\beqa
U(G)(c_\alpha, c^{\dag }_\alpha ,\frac{1}{\sqrt{2}}) (-1)^n U^{\dag }(G)
=
(c_\beta, c^{\dag }_\beta, \frac{1}{\sqrt{2}}) (-1)^n
\left[ 
\BA{ccc} 
A_{\beta \alpha } & \overline{B}_{\beta \alpha } & 
{\displaystyle -\frac{\overline{x}_\beta }{\sqrt{2}}} \\
B_{\beta \alpha } & \overline{A}_{\beta \alpha } & 
{\displaystyle \frac{x_\beta }{\sqrt{2}}} \\
{\displaystyle \frac{y_\alpha }{\sqrt{2}}} & 
{\displaystyle -\frac{\overline{y}_\alpha }{\sqrt{2}}} & z 
\EA 
\right] ,
\label{SO2Nplus1chiraltrans}
\eeqa
\vspace{-0.7cm}
\beqa
\left.
\BA{ll}
&
A_{\alpha \beta }
\equiv
a_{\alpha \beta }
-
\overline{X}_\alpha Y_\beta
=
a_{\alpha \beta }
-
{\displaystyle \frac{\overline{x}_\alpha y_\beta }{2(1 + z)}} ,~~
B_{\alpha \beta }
\equiv
b_{\alpha \beta }
+
X_\alpha Y_\beta
=
b_{\alpha \beta }
+
{\displaystyle \frac{x_\alpha y_\beta }{2(1 + z)}} ,\\
\\[-8pt]
&
x_\alpha
\equiv
2ZX_\alpha ,~
y_\alpha
\equiv
2ZY_\alpha ,~
z
\equiv
2Z^2 - 1 .
\EA \!\!
\right\}
\label{relAtoaXY}
\eeqa\\[-10pt]
By using the relation 
$
U(G)(c, c^{\dag },\frac{1}{\sqrt{2}}) U^{\dag }(G)
=
U(G)(c, c^{\dag },\frac{1}{\sqrt{2}}) U^{\dag }(G)
(z + \rho)(-1)^n
$ 
and
the third column equation of
(\ref{SO2Nplus1chiraltrans}),
Eq.
(\ref{SO2Nplus1chiraltrans})
can be written as\\[-10pt]
\beq
U(G)(c, c^{\dag },\frac{1}{\sqrt{2}}) U^{\dag }(G)
=
(c, c^{\dag }, \frac{1}{\sqrt{2}}) 
(z - \rho)G ,
\eeq
\beq
G 
\stackrel{\mathrm{def}}{=}
\left[ 
\BA{ccc} 
A & \overline{B} & {\displaystyle -\frac{\overline{x}}{\sqrt{2}}} \\
B & \overline{A} & {\displaystyle \frac{x}{\sqrt{2}}} \\
{\displaystyle \frac{y}{\sqrt{2}}} & 
{\displaystyle -\frac{\overline{y}}{\sqrt{2}}} & z 
\EA 
\right],
~~~~
G^{\dag }G = GG^{\dag } = 1_{2N+1} ,~~~
\det G
=
1 ,
\label{defG2}
\eeq
\beq
U\!(G)U\!(G') \!=\! U\!(GG') ,~
U\!(G^{\!-1}) \!=\! U^{\!-1}\!(G) \!=\! U^{\!\dag }\!(G) ,~
U\!(1_{\!2N\!+\!1}) \!=\! \mathbb{I}_G~(\mbox{unit operator on}~G) .
\eeq
($c$, $c^{\dag }$, $\frac{1}{\sqrt{2}}$) is 
a $(2N \!+\!1)$-dimensional row vector 
(($c_{\alpha }$), ($c^{\dag }_{\alpha }$), $\frac{1}{\sqrt{2}}$). 
$A \!=\! (A^{\alpha }_{~\beta })$ and $B \!=\! (B_{\alpha \beta })$ 
are $N \!\times\! N$ matrices. 
The $U(G)$ is a nonlinear transformation 
with a $q$-number gauge factor $z \!-\! \rho$, 
${\rho } 
\!\equiv\! 
x_{\alpha }c^{\dag }_{\alpha } \!-\! \overline{x}_{\alpha }c_{\alpha }$ and 
${\rho }^{2} 
\!=\! 
- \overline{x}_{\alpha }x_{\alpha } \!=\! {z}^{2} \!-\! 1$ 
\cite{Fuku.Yama.Nishi.77}.
The matrix $G$ is a matrix belonging to the $SO(2N \!\!+\!\! 1)$ group,
which is transformed to a real $(2N \!\!+\!\! 1)$-dimensional orthogonal matrix
as\\[-10pt]
\beq
O = VGV^{-1} ,~~
V
= 
\left[ 
\BA{ccc} 
{\displaystyle \frac{1}{\sqrt{2}} \!\cdot\! 1_N}&
{\displaystyle \frac{1}{\sqrt{2}} \!\cdot\! 1_N}&0 \\
-{\displaystyle \frac{i}{\sqrt{2}} \!\cdot\! 1_N}&
 {\displaystyle \frac{i}{\sqrt{2}} \!\cdot\! 1_N}&0
\\
0&0&1
\EA
\right] .
\eeq

The $SO(2N \!+\! 1)$ WF $\ket G \!=\! U(G) \ket 0$
\cite{Fuku.81,Fuku.77}
is expressed as\\[-20pt]
\beqa
\BA{l}
\ket G
\!=\!
\bra 0 U(G) \ket 0 (1 + r_\alpha c^\dagger_\alpha)
\exp({\displaystyle \frac{1}{2}} 
\cdot q_{\alpha\beta }c_\alpha^\dagger c_\beta^\dagger) 
\ket0 ,~~
r_\alpha
\!\equiv\!
{\displaystyle \frac{1}{1+z}}
(x_\alpha + q_{\alpha \beta } \overline{x}_\beta) ,
\EA
\label{SO2Nplus1wf}
\eeqa
\vspace{-1.0cm}
\beqa
\bra0\, U(G)\,\ket0 
= 
\overline{\Phi }_{00}(G) 
= 
\sqrt{\frac{1+z}{2}}
\left[
\det(1_N + q^\dag q)
\right]^{-\frac{1}{4}}
e^{i\frac{\tau }{2}} .
\label{SO2Nplus1vacuumf}
\eeqa\\[-32pt]

Following Fukutome
\cite{Fuku.81},
we define the projection operators $P_{\!+}$ and $P_{\!-}$ 
onto the sub-spaces of
even and odd fermion numbers, respectively, by\\[-12pt]
\beq
P_\pm
\stackrel{\mathrm{def}}{=}
{\displaystyle \frac{1}{2}}(1 \pm (-1)^n) ,~~
P_\pm ^2
=
P_\pm ,~~
P_+ P_-
=
0 ,
\eeq\\[-16pt]
and define the following operators with the supurious index 0:\\[-20pt]
\beqa
\left.
\BA{ll}
&E^0_{~0}
\stackrel{\mathrm{def}}{=}
-
{\displaystyle \frac{1}{2}} (-1)^n
=
{\displaystyle \frac{1}{2}} (P_- - P_+) ,\\
\\[-12pt]
&E^\alpha_{~0}
\stackrel{\mathrm{def}}{=}
c^\dagger_\alpha P_- 
=
P_+ c^\dagger_\alpha,~~
E^0_{~\alpha }
\stackrel{\mathrm{def}}{=}
c_\alpha P_+ 
=
P_-c_\alpha ,\\
\\[-12pt]
&E^{\alpha 0}
\stackrel{\mathrm{def}}{=}
-c_\alpha^\dagger P_+ 
=
-P_- c_\alpha^\dagger,~~
E^{0 \alpha }
\stackrel{\mathrm{def}}{=}
-E^{\alpha 0} ,\\
\\[-12pt]
&E_{\alpha 0}
\stackrel{\mathrm{def}}{=}
c_\alpha P_- 
=
P_+ c_\alpha ,~~
E_{0 \alpha }
\stackrel{\mathrm{def}}{=}
-E_{\alpha 0} .
\EA
\right\}
\label{spuriousoperators}
\eeqa\\[-18pt]
The annihilation-creation operators can be expressed 
in terms of the operators
(\ref{spuriousoperators})
as\\[-12pt]
\beq
c_\alpha
=
E_{\alpha 0} + E^0_{~\alpha },~~
c^\dag_\alpha
=
-E^{\alpha 0} + E^\alpha_{~0} .
\eeq\\[-16pt]
We introduce the indices $p, q, \cdots$ running over  
$(N \!\!+\!\! 1)$ values $0,1,\cdots,N$.
Then the operators of 
(\ref{operatorset}) and (\ref{spuriousoperators})
can be denoted in a unified manner as
$E^p_{~q},~E_{pq}$ and $E^{pq}$.
They satisfy\\[-18pt]
\beqa
\left.
\BA{ll}
&
E^{p \dag }_{~q}
=
E^q_{~p} ,~~
E^{p q } 
=
E^{\dag }_{q p } ,~~
E_{p q }
=
- E_{q p } ,~~
(p , q = 0, 1, \cdots,~N) \\
\\[-8pt]
&
[E^p_{~q},~E^r_{~s}]
=
\delta_{q r}E^p_{~s} 
- 
\delta_{p s}E^r_{~q},~~
(U(N+1)~\mbox{algebra}) \\
\\[-8pt]
&
\left.
\BA{ll}
&[E^p_{~q},~E_{r s}]
=
\delta_{p s}E_{q r} 
- 
\delta_{p r}E_{q s},~~[E_{pq},~E_{rs}]
=
0 , \\
\\[-8pt]
&[E^{p q},~E_{r s}]
=
\delta_{p s}E^q_{~r}
+ 
\delta_{q r}E^p_{~s}
-
\delta_{p r}E^q_{~s} 
- 
\delta_{q s}E^p_{~r} .
\EA
\right]
\EA
\right\}
\label{commurelp}
\eeqa\\[-10pt]
The above commutation relations in
(\ref{commurelp}) 
are of the same form as 
(\ref{commurel1}) and (\ref{commurel2}).

Instead of 
(\ref{spuriousoperators}),
it is possible to employ the operators\\[-16pt]
\beqa
\tilde{E}^0_{~0}
=
{\displaystyle \frac{1}{2}} (-1)^n
=
{\displaystyle \frac{1}{2}} (P_+ - P_-) ,~~
\tilde{E}^\alpha_{~0}
=
c^\dagger_\alpha P_+ ,~~
\tilde{E}^0_{~\alpha }
=
c_\alpha P_- .
\label{spuriousoperators2}
\eeqa
Denoting
$E^\alpha_{~\beta } \equiv \tilde{E}^\alpha_{~\beta }$,
it is shown that the operators 
$\tilde{E}^p_{~q},~p,~q = 0,~1,~\cdots,~N$,
satisfy\\[-8pt]
\beq
\tilde{E}^{p \dag }_{~q}
=
\tilde{E}^q_{~p},~~
[\tilde{E}^p_{~q},~\tilde{E}^r_{~s}]
=
\delta_{q r}\tilde{E}^p_{~s} 
- 
\delta_{p s}\tilde{E}^r_{~q}.~~
(\tilde{U}\!(N \!+\! 1)~\mbox{algebra})
\label{commurel1tilde}
\eeq
The Lie algebra $\tilde{U}\!(N \!+\! 1)$ 
is a 
$U\!(N \!+\! 1)$ Lie algebra but
it is not unitarily equivalent to 
$U\!(N \!+\! 1)$.

The $SO(2N \!\!+\!\! 1)$ group is embedded into an $SO(2N \!\!+\!\! 2)$ group. 
The embedding leads us to an unified formulation of the $SO(2N \!\!+\!\! 1)$
regular representation in which paired and unpaired modes are
treated in an equal way. 
Define 
$(N \!\!+\!\! 1) \!\times\! (N \!\!+\!\! 1)$ matrices ${\cal A}$ and ${\cal B}$ as\\[-10pt]  
\beq
{\cal A}
= 
\left[ \!
\BA{cc}
A & {\displaystyle -\frac{\overline{x}}{2}}  \\
\\[-14pt]
{\displaystyle \frac{y}{2}} & {\displaystyle \frac{1+z}{2}}
\EA \!
\right],
~~~
{\cal B}
= 
\left[ \!
\BA{cc}
B & {\displaystyle \frac{x}{2}}  \\
\\[-14pt]
{\displaystyle -\frac{y}{2}} & {\displaystyle \frac{1-z}{2}}
\EA \!
\right],~~~
y = x^{\mbox{\scriptsize T}} \! a - x^{\dag }b.
\label{calAcalB}
\eeq
Imposing the ortho-normalization of the $G$, 
matrices ${\cal A}$ and ${\cal B}$ 
satisfy the ortho-normalization condition and then form an $SO(2N \!+\! 2)$ 
matrix ${\cal G}$
\cite{Fuku.77}
represented as\\[-10pt]
\beq
{\cal G}
= 
\left[ \!
\BA{cc}
{\cal A} & \overline{\cal B} \\
{\cal B} & \overline{\cal A}
\EA \!
\right],               
~~~~
{\cal G}^{\dag } {\cal G}
=
{\cal G}{\cal G}^{\dag }
= 1_{2N+2} ,
\label{calG}
\eeq\\[-10pt]
which means
the ortho-normalization conditions of the $N \!+\! 1$-dimensional
HB amplitudes\\[-16pt]
\beqa
\left.
\BA{ll}
&
{\cal A}^\dag {\cal A} + {\cal B}^\dag {\cal B} 
= 
1_{N+1} ,~~
{\cal A}^{\mbox{\scriptsize T}}{\cal B} 
+ 
{\cal B}^{\mbox{\scriptsize T}}{\cal A} 
= 
0 ,\\
\\[-10pt]
&
{\cal A} {\cal A}^\dag 
+ 
\overline{\cal B} {\cal B}^{\mbox{\scriptsize T}} 
= 
1_{N+1} ,~~
\overline{\cal A} {\cal B}^{\mbox{\scriptsize T}} 
+ 
{\cal B}{\cal A}^\dag 
= 
0 .
\EA
\right\}
\label{HBorthonormalization}
\eeqa\\[-12pt]
The matrix ${\cal G}$ satisfies $\det {\cal G} = 1$ 
as is proved easily below\\[-12pt]
\beq
\det {\cal G}
=
\det \!
\left( \!
{\cal A} - \overline{\cal B}~\! \overline{\cal A}^{-1} {\cal B} \!
\right)
\det \overline{\cal A}
=
\det \!
\left( \!
{\cal A}{\cal A}^\dag 
- 
\overline{\cal B}~\! \overline{\cal A}^{-1} {\cal B}{\cal A}^\dag \!
\right)
\!=\! 1 .
\label{detcalG}
\eeq\\[-16pt]
By using
(\ref{relAtoaXY}) and (\ref{defG}),
the matrices ${\cal A}$ and ${\cal B}$
can be decomposed as\\[-10pt]  
\beq
{\cal A}
\!=\! 
\left[ 
\BA{cc}
1_N \!-\! {\displaystyle \frac{\overline{x} r^{\mbox{\scriptsize T}}}{2}} & 
{\displaystyle -\frac{\overline{x}}{2}} \\
\\[-12pt]
{\displaystyle \frac{(1\!+\!z)r^{\mbox{\scriptsize T}}}{2}} & 
{\displaystyle \frac{1\!+\!z}{2}}
\EA 
\right] \!\!
\left[ 
\BA{cc}
a & 0 \\
\\ \\[-4pt]
0 & 1
\EA 
\right],
~
{\cal B}
\!=\! 
\left[ 
\BA{cc}
1_N \!+\! {\displaystyle \frac{x r^{\mbox{\scriptsize T}}q^{-1}}{2}} & 
{\displaystyle \frac{x}{2}} \\
\\[-12pt]
- {\displaystyle \frac{(1\!+\!z)r^{\mbox{\scriptsize T}}q^{-1}}{2}} & 
{\displaystyle \frac{1\!-\!z}{2}}
\EA 
\right] \!\!
\left[ 
\BA{cc}
b & 0 \\
\\ \\[-4pt]
0 & 1
\EA 
\right] ,
\label{calApcalBp}
\eeq
from which we get
the inverse of ${\cal A},~{\cal A}^{-1}$ and a $\frac{SO(2N \!+\! 2)}{U(N \!+\! 1)}$
coset variable ${\cal Q}$ as\\[-10pt] 
\beq
{\cal A}^{-1}
=
\left[ \!
\BA{cc}
a^{-1} & 0 \\
\\[-4pt]
0 & 1
\EA \!
\right] \!\!
\left[ \!
\BA{cc}
1_N & {\displaystyle \frac{\overline{x}}{1\!+\!z}} \\
\\[-14pt]
- r^{\mbox{\scriptsize T}} & 1
\EA \!
\right] ,~
{\cal Q}
=
{\cal B}{\cal A}^{-1}
\!=\! 
\left[ \!
\BA{cc}
q & r \\
\\[-12pt]
-r^{\mbox{\scriptsize T}} & 0
\EA \!
\right]
=
-{\cal Q}^{\mbox{\scriptsize T}}.
\label{calAinversecosetvariable}
\eeq\\[-10pt]
The variables 
$q_{\alpha \beta }$ and $r_\alpha$
are independent variables of the
$\frac{SO(2N \!+\! 2)}{U(N \!+\! 1)}$ coset space. 
The paired mode $q_{\alpha \beta }$ and 
unpaired mode $r_\alpha$ variables in the $SO(2N \!\!+\!\! 1)$ algebra 
are unified as the paired variables in the $SO(2N \!+\! 2)$ algebra$\!$
\cite{Fuku.77}.$\!$
We denote the $(N \!+\! 1)$-dimension of the matrix $Q$ by
the index 0 and use the indices $p,~q,~\cdots$ 
running over 0 and $\alpha,~\beta,~\cdots$.
Using the $(2N \!+\! 2) \!\times\! (N \!+\! 1)$ isometric matrix 
$
{\cal U}
({\cal U}^{\mbox{\scriptsize T}}
\!\!=\!\!
[{\cal B}^{\mbox{\scriptsize T}}, {\cal A}^{\mbox{\scriptsize T}}],
{\cal U}^\dag{\cal U} \!\!=\!\! 1_{N \!+\! 1})
$,
let us introduce the $(2N \!+\! 2) \!\times\! (2N \!+\! 2)$ matrix
${\cal W}$:\\[-10pt]
\beq
{\cal W}
=
{\cal U}{\cal U}^\dag
=
\left[ \!
\BA{cc} 
{\cal R} & {\cal K} \\
\\[-8pt]
-\overline{\cal K} & 1_{N+1} - \overline{\cal R}
\EA 
\right] ,~~
\BA{c}
{\cal R}
=
{\cal B}{\cal B}^\dag ,\\
\\[-10pt]
{\cal K}
=
{\cal B}{\cal A}^\dag  ,
\EA \!
\label{densitymat}
\eeq\\[-8pt]
which satisfies the idempotency relation 
${\cal W}^2 \!=\! {\cal W}$
and is hermitian
on the $SO(2N \!+\! 2)$ group.
The ${\cal W}$ is a natural extension of 
the generalized density matrix in the $SO(2N)$ coherent state (CS) rep
to that in the $SO(2N \!+\! 2)$ CS rep.
As was shown also in Ref.$\!$
\cite{Nishi.Provi.08},
both the matrices ${\cal A}$ and ${\cal B}$ and
both the matrices ${\cal R}$ and ${\cal K}$
are represented in terms of only ${\cal Q}\!\!=\!\!({\cal Q}_{\!pq})$ as\\[-10pt]
\beq
{\cal A}
=
(1_{N+1} + {\cal Q}^\dag{\cal Q})^{-\frac{1}{2}}
\stackrel{\circ}{{\cal U}} ,~~
{\cal B}
=
{\cal Q}
(1_{N+1} + {\cal Q}^\dag{\cal Q})^{-\frac{1}{2}}
\stackrel{\circ}{{\cal U}} ,~~
\stackrel{\circ}{{\cal U}} \in U(N+1) ,
\label{matAandB}
\eeq
\vspace{-0.4cm}
\beq
{\cal R}
=
{\cal Q}(1_{N+1} + {\cal Q}^\dag{\cal Q})^{-1}{\cal Q}^\dag
\!=\!
1_{\!N\!+\!1}
\!-\!
(1_{\!N\!+\!1} \!+\! {\cal Q}{\cal Q}^\dag)^{\!-1} ,~~
{\cal K}
=
{\cal Q}
(1_{N+1} + {\cal Q}^\dag{\cal Q})^{-1} .
\label{matRandK}
\eeq

Finally, the Hamiltonian of the fermion system
under consideration is given as\\[-16pt]
\ba
H
\!=\!
h_{\alpha\beta } \!
\left( \! E^\alpha_{~\beta } \!+\! \frac{1}{2}\delta_{\alpha\beta } \! \right)
\!+\!
\frac{1}{4}[\alpha\beta|\gamma\delta]
E^{\alpha\gamma }E_{\delta\beta } ,
\label{Hamiltonian}
\ea\\[-12pt]
in which $h_{\alpha\beta }$ is a single-particle hamiltonian with a chemical
potential and
$
[\alpha\beta|\gamma\delta]
\!\!=\!\!
-
[\alpha\delta|\gamma\beta]
\!\!=\!\!
[\gamma\delta|\alpha\beta]
\!\!=\!\!
\overline{[\beta\alpha|\delta\gamma]}
$
are anti-symmetrized matrix
elements of an interaction potential.


\newpage

 
\setcounter{equation}{0}
\renewcommand{\theequation}{\arabic{section}.\arabic{equation}}

\section{$\!\!$Differential form for boson 
over $\frac{\mbox{\bf SO(2N\!+\!2)}}{\mbox{\bf U(N\!+\!1)}}$ coset space}


\vspace{-0.2cm}
 
~~~
The boson images
\mbox{\boldmath ${\cal E}^p_{~q}$}$~\!$etc. of the operators
$E^p_{~q}~\!$etc.  in
(\ref{commurelp})
are represented by the closed first order differential forms
over the $\frac{SO(2N \!+\! 2)}{U(N \!+\! 1)}$ coset space
in terms of the $\frac{SO(2N \!+\! 2)}{U(N \!+\! 1)}$ coset variables ${\cal Q}_{pq}$ 
and 
the phase variable
$
\tau \!
\left(
\!=\!
\frac{i}{2} \ln \!
\left[\frac{\det(\overline{A})}{\det({A})}
\right] \!
\right)
$  
of the $U(N \!+\! 1)$,
being identical with that of the $U(N)$,
$
\tau \!
\left(
\!=\!
\frac{i}{2} \ln \!
\left[\frac{\det(\overline{a})}{\det({a})}
\right] \!
\right)
$, 
due to the relation
$\det {\cal A} \!=\! \frac{1+z}{2} \det a$, 
in the following forms:\\[-12pt]
\beqa
\BA{c}
\mbox{\boldmath ${\cal E}^p_{~q}$} 
\!=\!
{\displaystyle
\overline{{\cal Q}}_{pr}\frac{\partial }{\partial \overline{{\cal Q}}_{qr}}
\!-\!
{\cal Q}_{qr}\frac{\partial }{\partial {\cal Q}_{pr}}
\!-\!
i\delta_{pq}\frac{\partial }{\partial \tau }
} ,~
\mbox{\boldmath ${\cal E}_{pq}$} 
\!=\!
{\displaystyle
{\cal Q}_{pr}{\cal Q}_{sq}\frac{\partial }{\partial {\cal Q}_{rs}}
\!-\!
\frac{\partial }{\partial \overline{{\cal Q}}_{pq}}
\!-\!
i{\cal Q}_{pq}\frac{\partial }{\partial \tau }
} ,~
\mbox{\boldmath ${\cal E}^{pq}$}
\!=\!
\mbox{\boldmath $\overline{\cal E}_{pq}$} ,
\EA
\label{SO2Nplus2LieopQ}
\eeqa\\[-10pt]
which are derived in a way quite analogous to 
the $SO(2N)$ case of the fermion Lie operators.
From 
(\ref{SO2Nplus2LieopQ}),
we can get the images of the fermion $SO(2N \!\!+\!\! 1)$ Lie operators,
namely,
the representations of the $SO(2N \!\!+\!\! 1)$ Lie operators 
in terms of
the variables $q_{\alpha \beta }$ and $r_\alpha$
\cite{Fuk.77}:\\[-14pt]
\beqa
\!\!\!\!\!\!\!\!
\left.
\BA{ll}
&\E^\alpha_{~\beta }
=
\mbox{\boldmath ${\cal E}^\alpha_{~\beta }$}
=
{\displaystyle
\mbox{\boldmath $e^\alpha_{~\beta }$}
+
\overline{r}_\alpha\frac{\partial }{\partial \overline{r}_\beta }
-
r_\beta\frac{\partial }{\partial r_\alpha } ,~
\mbox{\boldmath $e^\alpha_{~\beta }$}
\!\equiv\!
\overline{q}_{\alpha \gamma }\frac{\partial }
{\partial \overline{q}_{\beta \gamma }}
-
q_{\beta \gamma }\frac{\partial }{\partial q_{\alpha \gamma }}
-
i\delta_{\alpha \beta }\frac{\partial }{\partial \tau }
}~,\\
\\[-16pt]
&\E_{\alpha \beta }
=
\mbox{\boldmath ${\cal E}_{\alpha \beta }$}
=
{\displaystyle
\mbox{\boldmath $e_{\alpha \beta }$}
+
(r_\alpha q_{\beta\xi}-r_\beta q_{\alpha \xi })
\frac{\partial }{\partial r_\xi } ,~
\mbox{\boldmath $e_{\alpha \beta }$}
\!\equiv\!
q_{\alpha \gamma }q_{\delta \beta }\frac{\partial }
{\partial q_{\gamma \delta }}
-
\frac{\partial }{\partial \overline{q}_{\alpha \beta }}
-
iq_{\alpha \beta }\frac{\partial }{\partial \tau }
}~,\\
\\[-14pt]
&\E^{\alpha \beta }
\!=\!
\mbox{\boldmath $\overline{E}_{\alpha \beta }$}
\!=\!
\mbox{\boldmath ${\cal E}^{\alpha \beta }$}
\!=\!
{\displaystyle
\mbox{\boldmath $e^{\alpha \beta }$}
\!+\!
(\overline{r}_\alpha \overline{q}_{\beta\xi}
\!-\!
\overline{r}_\beta \overline{q}_{\alpha \xi })
\frac{\partial }{\partial \bar{r}_\xi } ,~
\mbox{\boldmath $e^{\alpha \beta }$}
\!\equiv\!
\mbox{\boldmath $\overline{e}_{\alpha \beta }$}
\!=\!
\overline{q}_{\alpha \gamma }\overline{q}_{\delta \beta }\frac{\partial }
{\partial \overline{q}_{\gamma \delta }}
\!-\!
\frac{\partial }{\partial q_{\alpha \beta }}
\!+\!
i\overline{q}_{\alpha \beta }\frac{\partial }{\partial \tau }
} ,
\EA \!\!\!
\right\}
\label{SO2Nplus1Lieopa}
\eeqa
\vspace{-0.3cm}
\beqa
\!\!\!\!\!\!\!\!
\left.
\BA{ll}
&\mbox{\boldmath $c_{\alpha }$}
\!=\!
\mbox{\boldmath ${\cal E}_{0 \alpha }$}
-
\mbox{\boldmath ${\cal E}^0_{~\alpha }$}
\!=\!
{\displaystyle
\frac{\partial }{\partial \overline{r}_\alpha }
+
\overline{r}_\xi \frac{\partial }{\partial \overline{q}_{\alpha \xi }}
+
(r_{\alpha }r_\xi -q_{\alpha \xi })\frac{\partial }{\partial r_\xi }
-
q_{\alpha\xi }r_\eta\frac{\partial }{\partial q_{\xi \eta }}
+
ir_{\alpha }\frac{\partial }{\partial \tau }
} , \\
\\[-14pt]
&\mbox{\boldmath $c^\dagger_{\alpha }$}
\!=\!
\mbox{\boldmath ${\cal E}_{0 \alpha }$}
-
\mbox{\boldmath ${\cal E}^0_{~\alpha }$}
\!=\!
-
\mbox{\boldmath $\overline{c}_{\alpha }$}
\!=\!
-
{\displaystyle
\frac{\partial }{\partial r_\alpha }
-
r_\xi \frac{\partial }{\partial q_{\alpha \xi }}
-
(\overline{r}_{\alpha }\overline{r}_\xi - \overline{q}_{\alpha \xi })
\frac{\partial }{\partial \overline{r}_\xi }
+
\bar{q}_{\alpha\xi }\overline{r}_\eta\frac{\partial }{\partial \overline{q}_{\xi \eta }}
-
i\overline{r}_{\alpha }\frac{\partial }{\partial \tau }
} .
\EA \!\!
\right\}
\label{SO2Nplus1Lieopb}
\eeqa\\[-10pt]
The vacuum function $\Phi_{00}(G)$
satisfies
$
\mbox{\boldmath $c_{\alpha }$} \Phi_{00}(G)
\!=\!
0
$ and
$
\mbox{\boldmath $c^{\dag }_{\alpha }$} \Phi_{00}(G)
\!=\!
\overline{r}_\alpha \Phi_{00}(G) .
$
\def\erwt#1{{<\!\!#1\!\!>_G}}

Operating
(\ref{SO2Nplus1Lieopa})
and
(\ref{SO2Nplus1Lieopb})
on the function $\chi(\overline{Q},\tau) \Phi_{00}(G)$,
we define the operators $\widetilde{\E}^\alpha_{~\beta }$ etc.
as
$
\widetilde{\E}\chi(\overline{Q},\tau) \Phi_{00}(G) 
=
\Phi_{00}(G) \widetilde{\E}\chi(\overline{Q},\tau) .
$
Then, we get\\[-14pt]
\beqa
\left.
\BA{ll}
&\widetilde{\E}^\alpha_{~\beta }
=
{\displaystyle
\mbox{\boldmath $\widetilde{e}^\alpha_{~\beta }$}
+
\overline{r}_\alpha \frac{\partial }{\partial \overline{r}_\beta } ,~~~
\mbox{\boldmath $\widetilde{e}^\alpha_{~\beta }$}
+
\frac{1}{2} \delta_{\alpha \beta }
=
\overline{q}_{\alpha \gamma }\frac{\partial }
{\partial \overline{q}_{\beta \gamma }}
} , \\
\\[-16pt]
&\widetilde{\E}_{\alpha \beta }
=
{\displaystyle
\mbox{\boldmath $\widetilde{e}_{\alpha \beta }$} ,~~~
\mbox{\boldmath $\widetilde{e}_{\alpha \beta }$}
=
-
\frac{\partial }{\partial \overline{q}_{\alpha \beta }}
}, \\
\\[-14pt]
&\widetilde{\E}^{\alpha \beta }
=
{\displaystyle
\mbox{\boldmath $\widetilde{e}^{\alpha \beta }$}
+
\left(
\overline{r}_\alpha \overline{q}_{\beta \xi }
-
\overline{r}_\beta \overline{q}_{\alpha \xi }
\right)
\frac{\partial }{\partial \overline{r}_\xi },~~~
\mbox{\boldmath $\widetilde{e}^{\alpha \beta }$}
=
\overline{q}_{\alpha \beta }
+
\overline{q}_{\alpha \gamma }
\overline{q}_{\delta \beta }
\frac{\partial }{\partial \overline{q}_{\gamma \delta }}
} ,
\EA 
\right\}
\label{SO2Nplus1Lieopa2}
\eeqa
\vspace{-0.3cm}
\beqa
\left.
\BA{ll}
&\mbox{\boldmath $\widetilde{c}_{\alpha }$}
\!=\!
{\displaystyle
\frac{\partial }{\partial \overline{r}_\alpha }
+
\overline{r}_\xi \frac{\partial }{\partial \overline{q}_{\alpha \xi }}
} ,~ \\
\\[-14pt]
&\mbox{\boldmath $\widetilde{c}^\dagger_{\alpha }$}
\!=\!
\overline{r}_\alpha
+
\left(
\overline{q}_{\alpha \xi }
-
\overline{r}_\alpha \overline{r}_\xi
\right)
{\displaystyle
\frac{\partial }{\partial \overline{r}_\xi }
+
\overline{q}_{\alpha \xi }
\overline{r}_{\eta }
\frac{\partial }{\partial \overline{q}_{\xi \eta }}
} .
\EA 
\right\}
\label{SO2Nplus1Lieopb2}
\eeqa\\[-8pt]
The $\chi(\overline{Q},\tau)$
is not necessarily antisymmetric and valid in a space wider than 
the $SO(2N \!+\! 1)$ spinor space.
Eq.
(\ref{SO2Nplus1Lieopa2})
is identical with the Dyson rep
\cite{Dyson.56,JDFJ.71,Okubo.74,Fuku.77} 
for the paired operators
if we regard
$
-
\frac{\partial }{\partial \overline{q}_{\alpha \beta }}
$
-
$-\overline{q}_{\alpha \beta }$
formally as boson annihilation-creation operators 
with double indices.
Eq.
(\ref{SO2Nplus1Lieopb2})
is also identical with the Dyson rep
for the unpaired operators,
regarding
$
-
\frac{\partial }{\partial \overline{r}_{\alpha }}
$
-
$-\overline{r}_{\alpha }$
formally as boson annihilation-creation operators
with a single index.
Equations
(\ref{SO2Nplus1Lieopa2})
and
(\ref{SO2Nplus1Lieopb2})
are exact extensions of the Dyson rep
to include paired and unpaired modes.
They provide a consistent boson formalism
covering even and odd fermion-number systems.


\newpage

\def\thesection{\arabic{section}}
\setcounter{equation}{0}
\renewcommand{\theequation}{\arabic{section}.
\arabic{equation}}
\section{Matrix-valued generator coordinate and modified non-Euclidian transformation}

\def\bra#1{{<\!#1\,|}} 
\def\ket#1{{|\,#1\!>}}

~~~Consider 
a fermion state vector $\ket \Psi$ 
corresponding to a function 
$\Psi ({\cal G})$ in ${\cal G} \!\in\! SO(2N\!+\!2)$, \\[-24pt] 
\beqa
\BA{l}
\ket \Psi
\!=\!\!
{\displaystyle \int} \! 
U ({\cal G}) \ket 0 \bra 0 U^\dag ({\cal G}) \ket \Psi d{\cal G}
\!=\!\!
{\displaystyle \int} \!
U ({\cal G}) \ket 0 \Psi ({\cal G}) d{\cal G} ,
\EA
\label{statePsi}
\eeqa\\[-20pt]  
and
a state $|f \rangle$,
an exact representaion on the $SO(2N \!\!+\!\! 1)$, \\[-24pt] 
\begin{eqnarray}
|f \rangle 
=\!\!~   
2^N \!\!\! \int \!\! U(G) | 0 \rangle 
\langle 0 | U^{\dagger} (G) |f \rangle dG   
=\!\!~   
2^N \!\!\! \int \! | G \rangle \Phi_{0f}(G) dG ,
\label{integralrep}    
\end{eqnarray}\\[-20pt]
where
$d{\cal G}$ and $dG$ are invariant group integrations over the
$\!SO(2N \!\!+\!\! 2)\!$ and $\!SO(2N \!\!+\!\! 1)\!$ groups.
From
(\ref{statePsi})
the invariance of group measure
of transformations by any group element
leads to\\[-24pt]
\begin{eqnarray}  
U(G) |f \rangle 
=\!\!~   
2^N \!\!\! \int \! | G' \rangle \Phi_{0f}(G^{\dagger}G') dG' ,
\label{integralrep2}    
\end{eqnarray}\\[-22pt]
which shows that
the canonical transformation to the $G$ frame
corresponds to a mere left coordinate transformation by $\!G^{\dagger}\!$ of
the matrix-valued generator coordinate $G'$.
Let $\cal G$ and $\cal G'$  be the $SO(2N \!\!+\!\! 2)$  matrices
corresponding to $G$ and $G'$, respectively.
The $\cal G$ is given by (\ref{calAcalB}) and (\ref{calG}).
Instead of $\cal G$,
let us introduce the matrix-valued generator coordinate $\widetilde{\cal G}$ 
in the $\cal G$ frame by  
$\widetilde{\cal G} \!=\! {\cal G}^{\dagger}{\cal G'}$.
Then, conversely, the $\cal G'$ is represented as\\[-22pt]    
\begin{eqnarray} 
\begin{array}{rl}
{\cal G'}
\!=\! 
\left[ \!\!
\begin{array}{cc}
{\cal A}'& \overline{\cal B}'\\
{\cal B}'& \overline{\cal A}'
\end{array} \!\!
\right]
\!\!\!  
&
=
{\cal G} \widetilde{\cal G}    
=
\left[ \!\!
\begin{array}{cc}
{\cal A}&\overline{\cal B}\\
\\[-10pt]
{\cal B}&\overline{\cal A}
\end{array} \!\!
\right] \!
\left[ \!
\begin{array}{cc}
\tilde{\cal A}&\overline{\widetilde{\cal B}}\\
\tilde{\cal B}&\overline{\widetilde{\cal A}}
\end{array} \!
\right] 
\!=\!
\left[ \!
\begin{array}{cc}
{\cal A} \widetilde{\cal A} \!+\! \overline{\cal B} \widetilde{\cal B}&
{\cal A} \overline{\widetilde{\cal B}} \!+\! \overline{\cal B} \overline{\widetilde{\cal A}}\\
\\[-12pt]
{\cal B} \widetilde{\cal A} \!+\! \overline{\cal A} \widetilde{\cal B}&
{\cal B} \overline{\widetilde{\cal B}} \!+\!  \overline{\cal A} \overline{\widetilde{\cal A}} 
\end{array} \!
\right] .
\end{array}
\label{coordinatecalGprime}
\end{eqnarray}\\[-12pt]
From
(\ref{coordinatecalGprime}) and the definition of $\widetilde{\cal Q}$ as  
$\widetilde{\cal Q} 
\!\equiv\! 
\widetilde{\cal B} \widetilde{\cal A}^{-1}$ in the coordinate $\tilde{\cal G}$,
we obtain the relations\\[-20pt]
\begin{eqnarray}
\left.
\begin{array}{rl}
{\cal A}' &\!\!\!
\!=\! 
{\cal A} \widetilde{\cal A} \!+\! \overline{\cal B} \widetilde{\cal B}
\!=\!
\left\{
{\cal A} \!+\! \overline{\cal B} \widetilde{\cal B} \widetilde{\cal A}^{-1}
\right\} \!
\widetilde{\cal A} 
\!=\! 
\left\{
{\cal A} \!+\! \overline{\cal B} \widetilde{\cal Q}
\right\} \!
\widetilde{\cal A}
\!=\! 
{\cal A}
\left\{
1_{\!N\!+\!1} \!+\! {\cal A}^{-1} \overline{\cal B} \widetilde{\cal Q}
\right\} \!
\widetilde{\cal A} , \\
\\[-10pt]
{\cal B}' &\!\!\!
\!=\! 
{\cal B}\widetilde{\cal A} \!+\! \overline{\cal A} \widetilde{\cal B}
\!=\!
\left\{
{\cal B} \!+\! \overline{\cal A} \widetilde{\cal B}\widetilde{\cal A}^{-1}
\right\} \!
\widetilde{\cal A} 
\!=\!
\left\{
{\cal B} \!+\! \overline{\cal A} \widetilde{\cal Q}
\right\} \!
\widetilde{\cal A} .
\end{array}
\right\}
\label{AandB}
\end{eqnarray}\\[-14pt]
Similarly, from ${\cal G}^{\dagger} {\cal G} \!\!=\!\! 1_{\!N\!+\!1}$
$\!
({\cal A}^{\dagger} {\cal A}
\!+\! 
{\cal B}^{\dagger} {\cal B}
\!\!=\!\!
1_{\!N\!+\!1})
$
and ${\cal Q}$  
$({\cal Q} \!\equiv\! {\cal B}{\cal A}^{-1})$
in the coordinate $\cal G$,
we have\\[-22pt]
\begin{eqnarray}
\begin{array}{c}
\overline{{\cal A}}
\!+\! 
({\cal B}{\cal A}^{-1} )^{\mbox{\scriptsize T}} \overline{{\cal B}}
\!=\!
\overline{{\cal A}}
\!-\! 
{\cal Q} \overline{{\cal B}}
\!=\!
 ({\cal A}^{\mbox{\scriptsize T}} )^{-1}.
\end{array}
\label{ApB}
\end{eqnarray}\\[-22pt]
Define a variable 
$\!{\cal Q}' 
\!\!\equiv\!\! 
{\cal B'} \! {\cal A'}^{-1}\!$ in the coordinate $\!{\cal G}'\!$.
An $SO(2N \!+\! 2)$ WF generated by a canonical transformation to 
${\cal G}'$ frame takes a function of 
the generator coordinate $\widetilde{{\cal G}}$:
$| {\cal G}' \rangle
\!=\!
U({\cal G} \widetilde{\cal G}) \ket 0$
owing to the relation ${\cal G}' \!=\! {\cal G} \widetilde{\cal G}$.
Using 
(\ref{AandB}) and (\ref{ApB}),
the variable ${\cal Q}'$ is written as\\[-22pt]
\begin{eqnarray}
\!\!\!\!
\begin{array}{ll}
{\cal Q}' 
\!=&\!\!\!\!
\left\{ \!
{\cal B} \!\!+\!\! \overline{\cal A} \widetilde{\cal Q} \!
\right\} \!\!
\left\{ \!
1_{\!N\!+\!1} \!\!+\!\! {\cal A}^{\!-1} \overline{\cal B} \widetilde{\cal Q} \!
\right\}^{\!-1} \!\!\!\!
{\cal A}^{\!-1} 
\!\!=\!\! 
\left[
{\cal B} \!
\left\{ \!
1_{\!N\!+\!1} \!\!+\!\! {\cal A}^{\!-1} \overline{\cal B} \widetilde{\cal Q} \!
\right\}
\!\!+\!\!
\left\{ \! - {\cal Q} \overline{\cal B}  
\!\!+\!\!
\overline{\cal A}
\right\} \!\! \widetilde{\cal Q}
\right] \!\!
\left\{ \!
1_{\!N\!+\!1} \!\!+\!\! {\cal A}^{-1} \overline{\cal B} \widetilde{\cal Q} \!
\right\}^{\!-1} \!\!\!\!
{\cal A}^{\!-1} \\
\\[-12pt]
~~~\!=  &\!\!\! 
{\cal Q}
\!+\! 
 ({\cal A}^{\mbox{\scriptsize T}} )^{-1}  \! \widetilde{\cal Q} 
\left\{
1_{\!N\!+\!1} \!+\! {\cal A}^{-1} \overline{\cal B} \widetilde{\cal Q}
\right\}^{-1} \!\!
{\cal A}^{-1} .
\end{array}
\label{noneuclidiantransformation0}
\end{eqnarray}\\[-18pt]
Let us introduce following matrices
${\cal R},~{\cal P}$ and ${\cal E}$:\\[-22pt]
\begin{eqnarray}
\begin{array}{l}
{\cal R} 
\!\equiv\!
- (\overline{\cal A})^{-1} {\cal B} , ~~ 
{\cal P} 
\!\equiv\! 
 ({\cal A}^{\mbox{\scriptsize T}} )^{-1} \! \widetilde{\cal Q} {\cal A}^{-1} , ~~
{\cal E}  
\!\equiv\! 
\overline{\cal A} {\cal B}^{\mbox{\scriptsize T}}
 \!=\!  
- {\cal B} {\cal A}^{\dagger}      
\!=\!  
- {\cal Q} (1_{\!N\!+\!1} \!-\! \overline{\cal Q} {\cal Q})^{-1}.
\end{array}
\label{matricesRPandE}
\end{eqnarray}\\[-24pt]
Then, the ${\cal Q}'$ is rewritten as\\[-24pt] 
\begin{eqnarray}
\begin{array}{l}
{\cal Q}'  
\!=\!
{\cal Q} 
\!+\!
{\cal P} \!
\left\{ \!
({\cal A} \!-\! {\cal A} \overline{\cal R} \widetilde{\cal Q}) 
{\cal A}^{-1} \!
\right\}^{-1} \!
\!=\!
{\cal Q} 
\!+\!
{\cal P}
(
1_{\!N\!+\!1} \!-\! {\cal A} \overline{\cal R} {\cal A}^{\mbox{\scriptsize T}}
{\cal P}
)^{-1} \!
\!=\!
{\cal Q}
\!+\!
{\cal P}
( 1_{\!N\!+\!1} \!-\! \overline{\cal E} {\cal P} )^{-1} ,
\end{array}
\label{noneuclidiantransformation}
\end{eqnarray}\\[-22pt]
whose transformation rule causes 
non-Euclidian properties of the coset variables 
because the coset variables are quantities 
defined on the non-commutative
$\frac{SO(2N+2)}{U(N+1)}$
Grassmann manifold.

Finally, we define the overlap integral of $SO(2N \!+\! 1)$ WFs\\[-22pt]
\begin{eqnarray}
\begin{array}{c}
S(G,G')
\!=\!
\overline{\Phi}_{00}(G^{\dagger} G') 
\!=\!
\langle 0 | U^{\dagger} (G) U(G') | 0 \rangle ,~
S(G,G)
\!=\!
1.
\end{array}
\label{overlapintegral}
\end{eqnarray}\\[-22pt]
Multiplying Eq. 
(\ref{integralrep})
by $\langle 0 | U^{\dagger} (G)$, 
we have\\[-22pt]
\begin{eqnarray}
\begin{array}{l}
\Phi_{0f} (G) 
\!=\! 
2^N \!\! {\displaystyle \int} \!
S(G,G') \Phi_{0f}(G') dG' ,
\end{array}
\label{fWF}
\end{eqnarray}\\[-16pt]
in which it is easily verified that
the overlap integral $S(G,G')$ satisfies\\[-24pt]
\begin{eqnarray}
\begin{array}{l}
S(G,G') 
\!=\! 
2^N \!\! {\displaystyle \int} \!  S(G,G'') S(G'',G') dG'' .
\end{array}
\label{projection operator}
\end{eqnarray}\\[-20pt]
This shows the $2^N \! S(G,G')$ is just the projection operator
to the $SO(2N \!+\! 1)$ spinor space.
Putting $\widetilde{G} \!\!=\!\! G^{\dagger}G'$ in
(\ref{overlapintegral})
and
using the same representations as those of
(\ref{AandB}),
we have\\[-22pt]
\begin{eqnarray}
\begin{array}{c}
\overline{\Phi}_{00}(G^{\dagger} G') 
=
\overline{\Phi}_{00}(\tilde{G}) 
=
\overline{\Phi}_{00}(\tilde{\cal G})
=
[\det(\tilde{\cal A})]^{\frac{1}{2}} ,
\end{array}
\label{0 state wave function}
\end{eqnarray}
\vspace{-0.9cm}
\begin{eqnarray}
\begin{array}{l}
\widetilde{\cal G} 
\!=\!
\left[ \!\!
\begin{array}{cc}
\widetilde{\cal A}& \overline{\widetilde {\cal B}}\\
\widetilde{\cal B}& \overline{\widetilde {\cal A}} 
\end{array} \!\!
\right]
\!= 
{\cal G}^{\dagger}{\cal G'}
\!=\!
\left[ \!\!
\begin{array}{cc}
{\cal A}^{\dagger}& {\cal B}^{\dagger} \\
\\[-12pt]
{\cal B}^{\mbox{\scriptsize T}}&{\cal A}^{\mbox{\scriptsize T}} 
\end{array} \!\!
\right] \!\!
\left[ \!\!
\begin{array}{cc}
{\cal A'}& \overline{\cal B'}\\
\\[-12pt]
{\cal B'}& \overline{\cal A'} 
\end{array} \!\!
\right] 
\!=\!
\left[ \!\!
\begin{array}{cc}
{\cal A}^{\dagger}{\cal A'} \!+\!  {\cal B}^{\dagger}{\cal B'}&
{\cal A}^{\dagger} \overline{\cal B'} \!+\!  {\cal B}^{\dagger} \overline{\cal A'} \\
\\[-12pt]
{\cal B}^{\mbox{\scriptsize T}}{\cal A'}
\!+\!
{\cal A}^{\mbox{\scriptsize T}}{\cal B'}&
{\cal B}^{\mbox{\scriptsize T}} \overline{\cal B'}
\!+\!
{\cal A}^{\mbox{\scriptsize T}} \overline{\cal B'}
\end{array} \!\!
\right] .
\end{array}
\label{coordinateGtilde}
\end{eqnarray}\\[-12pt]
Then, an explicit expression for the overlap integral
is obtained as\\[-22pt]
\begin{eqnarray}
\!\!\!\!\!\!\!\!
\begin{array}{ll}
S\!(\!G,G'\!) 
\!\!=\!\!
[\det \! (\tilde{\cal A})]^{\frac{1}{2}}
\!\!=\!\! 
\det ( \!\! {\cal A}^{\dagger} \! {\cal A'} \!\!+\!\! {\cal B}^{\dagger} \! {\cal B'} \! )
\!\!=\!\! 
D \!\! \left( \! {\cal Q}' \overline{\cal Q} \right) \!\!
\Phi_{00}(G) \overline{\Phi}_{00}(G') , 
D \!\! \left( \! {\cal Q}' \overline{\cal Q} \right) 
\!\!\equiv\!\!  
\left[ \! \det \!\! \left( 1_{\!N\!+\!1} \!\!-\!\! {\cal Q}' \overline{\cal Q} \right) \! \right]^{\!\frac{1}{2}} \!\! . 
\end{array}
\label{expression for S}
\end{eqnarray}\\[-32pt]

Taking the coordinate ${\cal G'}$
instead of the generator coordinate $\widetilde{\cal G}$
in 
(\ref{0 state wave function}), 
we have\\[-22pt]
\begin{eqnarray}
\begin{array}{c}
\Phi_{00}({\cal G'}) 
=
\langle 0 |  U^{\dagger}({\cal G'}) | 0 \rangle
= 
[\det(\overline{\cal A'})]^{\frac{1}{2}},~~
{\cal G'}
=
{\cal G} \widetilde{\cal G} .
\end{array}
\label{Phi(00)(g')}
\end{eqnarray}\\[-22pt]
Through
(\ref{AandB}) and (\ref{matricesRPandE}),
computation of a determinant of $\overline{\cal A'}$ is carried out as\\[-16pt]
\begin{eqnarray}
\!\!\!\!\!\!
\begin{array}{ll}
[\det(\overline{\cal A'})]^{\frac{1}{2}}
\!\!\!&\!=\!
[\det(\overline{\cal A})]^{\frac{1}{2}}
[\det(\overline{\tilde{\cal A}})]^{\frac{1}{2}} 
[\det (\overline{1_{\!N\!+\!1} \!+\! {\cal A}^{-1} \overline{\cal B} \tilde{\cal Q}})]^{\frac{1}{2}} \\
\\[-8pt]
&\!=\!
\Phi_{00}({\cal G}) \Phi_{00}(\tilde{\cal G}) 
[\det (
\overline{{\cal A}^{-1} \! {\cal A} \!\!+\!\! {\cal A}^{-1} \overline{\cal B}
{\cal A}^{\mbox{\scriptsize T}}{\cal P}{\cal A}}
)]
^{\frac{1}{2}}
\!\!=\!\!
\Phi_{00}({\cal G}) \Phi_{00}(\tilde{\cal G}) 
[\det (1_{\!N\!+\!1} \!-\! {\cal E} \overline{\cal P})]^{\frac{1}{2}}  .
\end{array}
\label{det of C'w'}
\end{eqnarray}\\[-12pt]
By using
(\ref{noneuclidiantransformation}),
the $SO(2N \!+\! 1)$ spinor function
$\Phi_{0f}(G')$
can be written as\\[-20pt]
\begin{eqnarray}
\begin{array}{ll}    
\Phi_{0f}(G') 
&\!\!\!=\! 
\chi_{\!f} (\overline{\cal Q}') \Phi_{00}(G') 
\!=\!
\chi_{\!f} \!
\left\{ \!
\overline{\cal Q} \!+\! \overline{P} \!
\left( 1_{\!N\!+\!1} \!-\! {\cal E} \overline{\cal P} \right)^{-1} \!
\right\} \! 
D ({\cal E} \overline{\cal P}) 
\Phi_{00}(G)  \Phi_{00}(\tilde{G}) , 
\end{array}
\label{Phi(0f)(g')1}
\end{eqnarray}\\[-18pt]
where
$\chi_{\!f} \! (\overline{\cal Q}')$
is an anti-symmetric polynomial of  $\overline{\cal Q}'_{pq}$
and
$D ({\cal E} \overline{\cal P})
\!\!\equiv\!\!
[\det (\! 1_{\!N \!+\! 1} \!-\! {\cal E} \overline{\cal P})]^{\frac{1}{2}}$.
We have used the non-Euclidian transformation
(\ref{noneuclidiantransformation})
and
(\ref{Phi(00)(g')}) and
(\ref{det of C'w'}).
From
(\ref{Phi(0f)(g')1}) 
we have\\[-20pt]
\begin{eqnarray}
\begin{array}{l}
\Phi_{0f}(G') 
\!=\! 
\chi_{\!f} (\overline{\cal Q} \!+\! K )
D ({\cal E} \overline{\cal P}) 
\Phi_{00}(G)  \Phi_{00}(\tilde{G}) ,~~
K
\!\equiv\!
\overline{\cal P}(1_{\!N\!+\!1} \!-\! {\cal E} \overline{\cal P} )^{-1} .
\end{array}
\label{Phi(0f)(g')2}
\end{eqnarray}\\[-20pt]
A differential formula for $D ({\cal E} \overline{\cal P})$
with respect to ${\cal E}_{pq}$
is easily given as\\[-18pt]
\begin{eqnarray}
\begin{array}{l}
{\displaystyle 
\frac{\partial D ({\cal E} \overline{\cal P}) }{\partial {\cal E}_{pq}}
=
\frac{\partial \det (1_{\!N\!+\!1} \!-\! {\cal E}\overline{\cal P} )}
{\partial (1_{\!N\!+\!1} \!-\! {\cal E} \overline{\cal P} )_{rs}}
\frac{\partial (1_{\!N\!+\!1} \!-\! {\cal E} \overline{\cal P} )_{rs}}{\partial {\cal E}_{pq}}
} 
=
K_{pq}   
D ({\cal E} \overline{\cal P})  , 
\end{array}
\label{differential formula}
\end{eqnarray}\\[-12pt]
where we have used the formula
$
\frac{\partial }{\partial {\cal A}^p_{~q}}
\det {\cal A}
\!=\!
({\cal A}^{-1})^{~q}_p
\det {\cal A}
$
for a regular matrix
$1_{\!N\!+\!1} \!-\! {\cal E} \bar{\cal P}
\!=\!
((1_{\!N\!+\!1} \!-\! {\cal E}\overline{\cal P} )_{rs})$.
As for the second differential for $D ({\cal E} \overline{\cal P}) $,
it is carried out as\\[-18pt]
\begin{eqnarray}
\!\!\!\!\!\!\!
\begin{array}{ll}
{\displaystyle 
\frac{\partial^2 \! D \!
\left( \! {\cal E} \overline{\cal P} \! \right) }
{\partial {\cal E}_{\!rs} \partial {\cal E}_{\!pq}}
} 
\!\!\!&\!\!=\!
{\displaystyle
\frac{\partial \! K_{\!qp}}{\partial {\cal E}_{\!rs}} \! D \!
\left( \! {\cal E} \overline{\cal P} \! \right)     
\!\!+\!\!
K_{\!sr} K_{\!qp} D \!
\left( \! {\cal E} \overline{\cal P} \! \right)  
}
\!\!=\!\!
\left[ \!
{\displaystyle
 \overline{\cal P}_{\!qp'}
 \frac{\partial \!
\left( \! 1_{\!N\!+\!1} \!\!-\!\! {\cal E} \overline{\cal P} \! \right)^{-1}_{p'p} }   
{\partial \!
\left( \! 1_{\!N\!+\!1} \!\!-\!\! {\cal E} \overline{\cal P} \! \right)_{uv}}
 \frac {\partial \!
\left( \! 1_{\!N\!+\!1} \!\!-\!\! {\cal E} \overline{\cal P} \! \right)_{\!uv}}{\partial {\cal E}_{\!rs}}
 }  
\!\!+\!\!
K_{\!sr} K_{\!qp} \! 
\right] \!\!
D \!
\left( \! {\cal E} \overline{\cal P} \! \right)  \\
\\[-10pt]
\!\!&\!\!=
\left[ K_{qp} K_{sr}  \!-\! K_{qr} K_{sp} \right] \!
D \!
\left( \! {\cal E} \overline{\cal P} \! \right)
\!=\!
\mathfrak{A}( K_{qp} K_{sr} )   
D \!
\left( \! {\cal E} \overline{\cal P} \! \right) ,
\end{array}
\end{eqnarray}\\[-12pt]
where
$\mathfrak{A}( K_{qp} K_{sr} )$
is the anti-symmetrized product of
$K_{qp} K_{sr}$.
Then,
successive differential calculation to higher orders lead to
a general differential formula\\[-10pt]
\begin{equation}
\frac{\partial^{l} \! D ({\cal E} \overline{\cal P})}
{\partial  {\cal E}_{pq} \cdots \partial  {\cal E}_{uv}}
\!=\!
\mathfrak{A}( K_{pq} \cdots  K_{uv})
D \! \left( {\cal E} \overline{\cal P} \right) .
\label{differential formulas}
\end{equation}\\[-12pt]
Applying the differential formulas of
$D ({\cal E} \overline{\cal P})$
with respect to ${\cal E}_{pq} \cdots {\cal E}_{uv}$,
the Tayler expansion of $\chi_{\!f} \! \left( \overline{\cal Q} \!+\! K \right)$ in 
(\ref{Phi(0f)(g')2})
with respect to $K$
is made as follows:\\[-18pt]
\begin{eqnarray}
\begin{array}{c}
\chi_{\!f} \!
\left( \! \overline{\cal Q} \!+\! {\displaystyle \frac{\partial}{\partial {\cal E}}} \! \right) \!
D \!
\left( \! {\cal E} \overline{\cal P} \! \right) 
\!=\!
\chi_{\!f} \!
\left( \overline{\cal Q} \right)  \! D \!
\left( \! {\cal E} \overline{\cal P} \! \right)   
\!+\!
\sum_{p < q} \!\!
{\displaystyle 
\frac{\partial \chi_{\!f} \!
\left( \overline{\cal Q} \right) }
{\partial  \overline{\cal Q}_{pq} }
\!\cdot\! 
\frac{\partial D \!
\left( \! {\cal E} \overline{\cal P} \! \right)}
{\partial {\cal E}_{pq}}
}
\!+\! 
\cdots \\
\\[-12pt]
\!+\!
\sum_{l=2, 3, \cdots} \! 
\sum_{p < \cdots < v} \!
{\displaystyle 
\frac{\partial^{l} \chi_{\!f} \!
\left( \overline{\cal Q} \right) }
{\partial  \overline{\cal Q}_{pq} \cdots \partial  \overline{\cal Q}_{uv} }
\!\cdot\! 
\frac{\partial^{l} \! D \!
\left( {\cal E} \overline{\cal P} \right)}
{\partial {\cal E}_{pq} \cdots \partial {\cal E}_{uv}}
}
\!+\!
\cdots .
\end{array}
\label{chifD(eq*)}
\end{eqnarray}\\[-20pt]

From now on we consider a fluctuation $ \widetilde{G}_0$ around
a stationary ground state $|G_0 \rangle$.
Due to the last relation of
(\ref{noneuclidiantransformation0}),
instead of
(\ref{noneuclidiantransformation0}),
we have
a modified non-Euclidian transformation\\[-16pt]
\begin{eqnarray}
\begin{array}{l}
{\cal Q}' 
\!=\!
{\cal Q}_0
+
 ({\cal A}_0^{\mbox{\scriptsize T}} )^{-1} \widetilde{\cal Q} 
\left\{ \!
1_{\!N\!+\!1}  - \overline{\widetilde{\cal Q}}_0 \widetilde{\cal Q}
\right\}^{\!-1} \!\!\!
{\cal A}_0^{-1} , ~
{\cal Q}_0 
\equiv 
{\cal B}_0 {\cal A}_0^{-1},~
\overline{\widetilde{\cal Q}}_0
\equiv
- {\cal A}_0^{-1} \overline{{\cal B}}_0 .
\end{array}
\label{noneuclidiantransformation1}
\end{eqnarray}\\[-12pt]
On the other hand, from
(\ref{matAandB}),
we have
the expressions for ${\cal A}_0$ and ${\cal B}_0$
as \\[-10pt]
\beq
{\cal A}_0
\!=\!
\left( 1_{\!N\!+\!1} - \overline{\cal Q}_0 {\cal Q}_0 \right)^{-\frac{1}{2}} \!
\stackrel{\circ}{\cal U}_0 ,~~
{\cal B}_0
\!=\!
{\cal Q}_0
\left( 1_{\!N\!+\!1} - \overline{\cal Q}_0 {\cal Q}_0 \right)^{-\frac{1}{2}} \!
\stackrel{\circ} {\cal U}_0 ,~~
\stackrel{\circ} {\cal U}_0
~\!\!=
1_{\!N\!+\!1} .
\label{matA0andB0}
\eeq\\[-16pt]
Thus, we obtain an important relation
$
\widetilde{\cal Q}_0
\!=\!
-
{\cal Q}_0 
$.
Let us carry modified quantities $\!\K\!$ and $\!\D\!$ given by
$
\K
\!\!=\!\!
\overline{\widetilde{\cal Q}} \!
\left( \!
1_{\!N\!+\!1} \!-\! \widetilde{\cal Q}_0  \overline{\widetilde{\cal Q}} 
\right)^{\!-1}
\!$  
and
$\!
\D \! \left(\! \widetilde{\cal Q}_0 \overline{\widetilde{\cal Q}} \!\right)
\!\!=\!\!
\left[ 
\det \!
\left( \!
1_{\!N\!+\!1} \!-\! \widetilde{\cal Q}_0 \overline{\widetilde{\cal Q}} 
\right) 
\right]^{\!\frac{1}{2}}
\!\!=\!\!
\D
$.
$\!$Then, we have the same types of the differential formulas as
(\ref{differential formula})
and
(\ref{differential formulas})
with respect to $\widetilde{\cal Q}_{0pq} \!\cdots\! \widetilde{\cal Q}_{0vw}$ as\\[-16pt]
\begin{eqnarray}
\!\!\!\!\!\!
\begin{array}{l}
{\displaystyle 
\frac{\partial \! \D }
{\partial \! \widetilde{\cal Q}_{0pq}}
}
\!=\!\!
\K_{\!pq}   
\D  , ~
{\displaystyle 
\frac{\partial^{n} \! \D }
{\partial  \! \widetilde{\cal Q}_{0pq} \!\cdots\! \partial \! \widetilde{\cal Q}_{0vw}}
}
\!=\!
\mathfrak{A}
( \! \K_{\!pq} \cdots  \K_{\!uv} \! ) \!
\D ,
\left( \!\!
\D
\!\!=\!\!
\sum_{\!n} \!\! \widetilde{\cal Q}_{0pq} \cdots \widetilde{\cal Q}_{0vw}
\mathfrak{A} \!
\left( \! \overline{\widetilde{\cal Q}}_{pq}
\!\cdots\!
\overline{\widetilde{\cal Q}}_{vw} \!
\right) \!
\right) \! .
\end{array}
\label{differential formula2}
\end{eqnarray}\\[-12pt]
We introduce a new differential operator
$ 
\frac{\partial }{\partial \overline{\cal E}_{pq}}
\!=\!
({\cal A}_0^{-1})^\dag_{pr} (\overline{\cal A}_0^{-1})_{sq}
\frac{\partial }{\partial \overline{\cal Q}_{rs}}
\!=\!
\{ 
({\cal A}_0^{-1})^\dag 
\frac{\partial }{\partial \overline{\cal Q}}
(\overline{\cal A}_0^{-1}) 
\}_{pq}
$.
The $\!SO(2N \!\!+\!\! 1)\!$ spinor function
$\!\chi \! (\overline{\cal Q}) \! \Phi_{00}(G)\!$
is given in the same way as
(\ref{Phi(0f)(g')1})
and written as\\[-12pt]
\begin{eqnarray}
\begin{array}{ll}    
&\chi (\overline{\cal Q}) \Phi_{00}(G) 
\!=\!
\Phi_{00}(\widetilde{G}) \Phi_{00}(G_0) 
\chi \!
\left\{
\overline{\cal Q}_0 + ({\cal A}_0^{-1} )^\dag \K ( \overline{\cal A}_0^{-1} ) 
\right\} \! 
\D \\
\\[-12pt]
&\!=\!
\Phi_{00}(\widetilde{G}) \Phi_{00}(G_0) 
\chi \!
\left\{
\overline{\cal Q}_0 + ({\cal A}_0^{-1} )^\dag
{\displaystyle \frac{\partial }{\partial \overline{\cal Q}_0}} 
( \overline{\cal A}_0^{-1} ) 
\right\} \!
\D \\
\\[-14pt]
&\!=\!
\Phi_{00}(\widetilde{G}) \Phi_{00}(G_0) \!
\left\{ \!
\chi \! \left( \overline{\cal Q}_0 \right) \! \D
\!+\! 
{\displaystyle \frac{\partial \chi} {\partial \overline{\cal E}_{pq}}} \! \left( \overline{\cal Q}_0 \right) \!
{\displaystyle \frac{\partial \D}{\partial \widetilde{\cal Q}_{0pq}}}
\!+\! 
{\displaystyle \frac{\partial^2 \chi} {\partial \overline{\cal E}_{pq}}} \! \left( \overline{\cal Q}_0 \right) \!
{\displaystyle \frac{\partial^2 \D}{\partial \widetilde{\cal Q}_{0pq}\partial\widetilde{\cal Q}_{0rs}}}
+
\cdots
\right\} \\
\\[-12pt]
&\!=\!
\Phi_{00}(\widetilde{G}) \Phi_{00}(G_0) \!
\left\{ {}^{^{^{^{^{^{}}}}}} \!\!
\chi \! \left( \overline{\cal Q}_0 \right) \!
\sum_n \!
\widetilde{\cal Q}_{0pq} \cdots \widetilde{\cal Q}_{0vw}
\mathfrak{A}\!
\left( \! \overline{\widetilde{\cal Q}}_{pq} \cdots  \overline{\widetilde{\cal Q}}_{vw} \! \right)
\right. \\
\\[-12pt]
&
\left.
\!+\!
{\displaystyle \frac{\partial \chi} {\partial \overline{\cal E}_{pq}}} \! \left( \overline{\cal Q}_0 \right) \!
\sum_n \!
\widetilde{\cal Q}_{0rs} \cdots \widetilde{\cal Q}_{0vw}
\mathfrak{A} \! \left( \! \overline{\widetilde{\cal Q}}_{pq} \overline{\widetilde{\cal Q}}_{rs}
\cdots  \overline{\widetilde{\cal Q}}_{vw} \! \right) 
\right. \\
\\[-12pt]
&
\left.
\!+\! 
{\displaystyle \frac{\partial^2 \chi} {\partial \overline{\cal E}_{pq} \partial \overline{\cal E}_{rs}}} \!
\left( \overline{\cal Q}_0 \right) \!
\sum_n \!
\widetilde{\cal Q}_{0tu} \cdots \widetilde{\cal Q}_{0vw}
\mathfrak{A} \!
\left( \! \overline{\widetilde{\cal Q}}_{pq} \overline{\widetilde{\cal Q}}_{rs}
\overline{\widetilde{\cal Q}}_{tu}
\cdots  \overline{\widetilde{\cal Q}}_{vw} \! \right)
+
\cdots
\right\} \\
\\[-12pt]
&\!=\!
\Phi_{00}(\widetilde{G}) \Phi_{00}(G_0) \!
\sum_n \!
\left\{ \!
\chi \! \left( \overline{\cal Q}_0 \right) \!
\widetilde{\cal Q}_{0pq} \cdots \widetilde{\cal Q}_{0vw}
\!+\!
{\displaystyle \frac{\partial \chi} {\partial \overline{\cal E}_{pq}}}\!
\left( \overline{\cal Q}_0 \right) \!
 \widetilde{\cal Q}_{0rs} \cdots \widetilde{\cal Q}_{0vw}
\right. \\
\\[-14pt]
&
\left.
\!+\! 
{\displaystyle \frac{\partial^2 \chi} {\partial \overline{\cal E}_{pq} \partial \overline{\cal E}_{rs}}} \!
\left( \overline{\cal Q}_0 \right) \!
\widetilde{\cal Q}_{0tu} \cdots \widetilde{\cal Q}_{0vw}
+
\cdots
\!+\! 
{\displaystyle \frac{\partial^n \chi} {\partial \overline{\cal E}_{pq} 
\cdots
\partial \overline{\cal E}_{vw}}}\!
\left( \overline{\cal Q}_0 \right) \!
\right\}
\mathfrak{A} \!
\left( \! \overline{\widetilde{\cal Q}}_{pq} \cdots  \overline{\widetilde{\cal Q}}_{vw} \! \right) \\
\\[-12pt]
&\!=\!
\Phi_{00}(G_0) \!\!
\sum_n \!\!
\left\{ \!
\chi \! \left( \overline{\cal Q}_0 \right) \!
{\cal E}_{0pq} \cdots {\cal E}_{0vw}
\!+\!
{\displaystyle \frac{\partial \chi} {\partial \overline{\cal E}_{pq}}} \!
\left( \overline{\cal Q}_0 \right) \!
{\cal E}_{0rs} \cdots{\cal E}_{0vw}
\!+\! 
{\displaystyle \frac{\partial^2 \chi} {\partial \overline{\cal E}_{pq} \partial \overline{\cal E}_{rs}}} \!
\left( \overline{\cal Q}_0 \right) \!
{\cal E}_{0tu} \cdots {\cal E}_{0vw}
\right. \\
\\[-12pt]
&
\left.
+
\cdots
\!+\! 
{\displaystyle \frac{\partial^n \chi} {\partial \overline{\cal E}_{pq} 
\cdots
\partial \overline{\cal E}_{vw}}} \!
\left( \overline{\cal Q}_0 \right) \!
\right\} \!
\mathfrak{A}
\left\{
\overline{[ ({\cal A}_0^{-1} )^{\mbox{\scriptsize T}}\widetilde{\cal Q}{\cal A}_0^{-1}]}_{pq}
\cdots 
\overline{[ ({\cal A}_0^{-1} )^{\mbox{\scriptsize T}}\widetilde{\cal Q}{\cal A}_0^{-1}]}_{vw} \!
\right\}
\Phi_{00}(\widetilde{G}) ,
\end{array}
\label{Phi(0f)(g')2}
\end{eqnarray}\\[-6pt]
where we have used the modified matrix in$\!\!$
(\ref{matricesRPandE})
$\!
{\cal E}_{\!0pq}  
\!\!\equiv\!\! 
[\overline{\cal A}_0 \! {\cal B}^{\mbox{\scriptsize T}}_0\!]_{pq}
$.
Introduction of a new operator
$
\Delta^n_{pq \cdots vw}
=
\mathfrak{A} 
\left\{ \!
{\cal E}_{0pq} \cdots {\cal E}_{0vw}
+
{\cal E}_{0rs} \cdots {\cal E}_{0vw}\frac{\partial} {\partial \overline{\cal Q}_{pq}} 
+
{\cal E}_{0tu} \cdots {\cal E}_{0vw}
\frac{\partial^2} {\partial \overline{\cal Q}_{pq} \overline{\cal Q}_{rs}}
+
\cdots
+
\frac{\partial^n} {\partial \overline{\cal Q}_{pq} \cdots \overline{\cal Q}_{rs}} \!
\right\}
$
leads to\\[-6pt] 
\begin{eqnarray}
\begin{array}{l}
\sum_l \!
\sum_{a<\cdots<g}
\Delta^k_{pq \cdots vw}
\overline{\Delta}^l_{ab \cdots fg} \!
\left\{
H^{ni}(\overline{\cal Q}, {\cal Q})
-
E^{ni}_{\lambda} S^{ni}(\overline{\cal Q}, {\cal Q}) 
\right\} \!
C^{nil}_{ab \cdots fg,\lambda}
\!=\!
0 ,
\end{array}
\label{FinalEq}
\end{eqnarray}\\[-8pt]
which is called the projected $SO(2N \!+\! 1)$
Tamm-Dancoff equation
whose original form has already been given in Ref.
\cite{Fuku.78}.
The
$E_{\lambda}$
means an eigenvalue.
The
$H(\overline{\cal Q}, {\cal Q})$
and
$S(\overline{\cal Q}, {\cal Q})$
are given soon later.
Finally, 
the eigenvectors
$C^{nil}_{ab \cdots fg,\lambda}$
are Tamm-Dancoff expansion coefficients.


\newpage


\def\thesection{\arabic{section}}
\setcounter{equation}{0}
\renewcommand{\theequation}{\arabic{section}.
\arabic{equation}}
\section{Classical TD SO(2N $\!+\!$ 1) Lagrangian}


\def\bra#1{{<\!#1\,|}} 
\def\ket#1{{|\,#1\!>}}

~~~
In this section we will derive a classical TD 
$SO(2N \!+\! 1)$ Lagrangian
describing collective excitations in even and odd Fermion systems,
respectively.
Following Ref.
\cite{Fuku.Nishi.84},
the Lagransian of the TD $SO(2N \!+\! 1)$ equation is given by\\[-18pt]
\begin{eqnarray}
{\cal L}
\!=\!
-
{\displaystyle
\frac{i \hbar}
{2}
} \!
\left( \!
<\!\Psi \,| \,\dot{\Psi} \!> 
-
<\!\dot{\Psi} \,| \,\Psi \!> \!
\right)
\!+\!
<\!\Psi \,| H |\,\Psi \!> , 
\label{Lagransian}
\end{eqnarray}\\[-20pt]
where the dot denotes time derivative.

$\!\!\!\!\!$Taking a coordinate ${\cal G'}$
instead of the generator coordinate $\widetilde{\cal G}$
in 
(\ref{0 state wave function}),
we use the $SO(2N \!\!+\!\! 1)$ WF
$
{\Phi}_{00}( \widetilde{G} G^\dag )
(\!=\!
{\Phi}_{00}(\widetilde{G}) \overline{\Phi}_{00}(G) 
D(Q \overline{\tilde{Q}})
)
$.
As basic preparations,
following Fukutome
\cite{pricommuF}, 
we calculate
the double group integration of the overlap integral
$
\Phi_{00}( \widetilde{G}') 
S(\overline{\widetilde{Q}'}, \widetilde{Q})
\overline{\Phi}_{00}( \widetilde{G})
$
over the $\widetilde{G}$ and $\widetilde{G}'$.
It is made in the following way as\\[-18pt]
\begin{eqnarray}
\begin{array}{l}
2^{2N} \!\!  
{\displaystyle \int} \!\!\!
{\displaystyle \int} \!
\overline{\Phi}_{00}(\widetilde{G}') 
D( \widetilde{Q}' \overline{Q} )
{\Phi}_{00}(\widetilde{G}')
S(\overline{\widetilde{Q}'}, \widetilde{Q})
\overline{\Phi}_{00}( \widetilde{G})
{\Phi}_{00}(\widetilde{G}) 
D(Q \overline{\widetilde{Q}} ) 
d \widetilde{G} d \widetilde{G}' \\
\\[-12pt]
\!\!\!=\!
2^{2N} \!\!
{\displaystyle \int} \!\!\!
{\displaystyle \int} \!
{\displaystyle
\frac{S(\overline{Q},  \widetilde{Q}')
\Phi_{00}(G)
\overline{\Phi}_{00}(\widetilde{G}')}
{\Phi_{00}(G)}
}     
S( \overline{\widetilde{Q}'}, \widetilde{Q} )
\Phi_{00}(\widetilde{G}') 
\overline{\Phi}_{00}( \widetilde{G}) 
{\displaystyle
\frac{S(\overline{\widetilde{Q}},Q) 
\Phi_{00}(\widetilde{G})
\overline{\Phi}_{00}(G) }
{\overline{\Phi}_{00}(G)}
} 
d \widetilde{G} d \widetilde{G}'  \\
\\[-12pt]
\!\!\!=\!
{\displaystyle
\frac{1}
{\Phi_{00}(G)}
} 
{\displaystyle
\frac{1}
{\overline{\Phi}_{00}(G)}
} 
2^N \!\! 
{\displaystyle \int} \!\! 
\left(  \!\!
2^N \!\! 
{\displaystyle \int} \! 
S(G,  \widetilde{G}')
S( \widetilde{G}', \widetilde{G} )
~\! d \widetilde{G}' \!
\right) \!
S(\widetilde{G}, G) 
d \widetilde{G} \\
\\[-12pt]
\!\!\!=\!
{\displaystyle
\frac{1}
{\left|\Phi_{00}(G)\right|^2 }
} 
2^N \!\! 
{\displaystyle \int} \! 
S(G,  \widetilde{G})
S(\widetilde{G}, G) 
d \widetilde{G} 
\!=\!
{\displaystyle
\frac{S(G, G) }
{\left|\Phi_{00}(G)\right|^2}
} 
\!=\!
S(\overline{Q}, Q)
\!=\!
\left[ \det \left( 1_{\!N\!+\!1} \!-\! {\cal Q} \overline{\cal Q} \right) \right]^{\!\frac{1}{2}} ,
\label{Phi(00)(tildeG')1}
\end{array}
\end{eqnarray}\\[-12pt]
where we have used the property of the projection operator
$2^N \! S(\!G, G'\!)\!$
(\ref{projection operator}).
The second equation in
(\ref{overlapintegral})
shows the relation
$S(G,G)\!=\!1$.
Using
(\ref{Phi(00)(tildeG')1}),
we also have\\[-20pt]
\begin{eqnarray}
\!\!\!\!\!\!
\begin{array}{l}
2^{2N} \!\!\!
{\displaystyle \int} \!\!\!
{\displaystyle \int} \!
\overline{\Phi}_{00}(\widetilde{G}' G^\dag ) \Phi_{00}( \widetilde{G}') 
S(\!\widetilde{G}', \!\widetilde{G}\!)
\overline{\Phi}_{00}( \widetilde{G})
\Phi_{00}(\widetilde{G} G^\dag )
d \widetilde{G} d \widetilde{G}' 
\!\!=\!\!
S(\!G, G\!) 
\!=\!
\left|\Phi_{00}(G)\right|^2 \!
S(\!\overline{Q}, Q\!)
\!=\!
1 .
\label{Phi(00)(tildeG')2}
\end{array}
\end{eqnarray}\\[-14pt]
Introducing the quantity $N$ defined as
$N \!\equiv\! |\Phi_{00}(G)| \! \left[S(\overline{Q}, Q)\right]^{\frac{1}{2}}$,
we obtain the relation\\[-18pt]
\begin{eqnarray}
\begin{array}{l}
{\displaystyle
\frac{\Phi_{00}(\widetilde{G} G^\dag )}{N}
}
\!=\!
{\Phi}_{00}(\widetilde{G}) 
D(Q \overline{\widetilde{Q}} ) \!
\left[ \!
{\displaystyle
\frac{\overline{\Phi}_{00}(G )}
{|\Phi_{00}(G)| }
} \!
\right] \!
\left[S(\overline{Q}, Q)\right]^{-\frac{1}{2}}  
\!=\!
{\Phi}_{00}(\widetilde{G}) 
D(Q \overline{\widetilde{Q}} )
e^{i\frac{\tau}{2}} \!
\left[S(\overline{Q}, Q)\right]^{-\frac{1}{2}} \! ,
\label{Phi(00)(tildeG')3}
\end{array}
\end{eqnarray}\\[-10pt]
owing to the form of expression for $\overline{\Phi}_{00}(\!G\! )$,$\!$
(\ref{SO2Nplus1vacuumf}).
Then, the first term of
(\ref{Lagransian}) is computed as\\[-16pt]
\begin{eqnarray}
\begin{array}{l}
-
{\displaystyle
\frac{i \hbar}
{2}
} \!
\left( \!
<\!\Psi \,| \,\dot{\Psi} \!> 
-
<\!\dot{\Psi} \,| \,\Psi \!> \!
\right) \\
\\[-12pt]
\!=\!
-
{\displaystyle
\frac{i \hbar}
{2}
} 
2^{2N} \!\!
{\displaystyle \int} \!\!\!
{\displaystyle \int} \!
S(\widetilde{G}', \widetilde{G}) 
\left\{ \!  
{\displaystyle
\frac{\overline{\Phi}_{00}(\widetilde{G}' G^\dag )}{N}
} 
{\displaystyle
\frac{\partial}{\partial t}
} \!\!
\left( \!\!
{\displaystyle
\frac{\Phi_{00}(\widetilde{G} G^\dag )}{N}
} \!\!
\right)
\!-\!
{\displaystyle
\frac{\partial}{\partial t}
} \!\!
\left( \!\!
{\displaystyle
\frac{\overline{\Phi}_{00}(\widetilde{G}' G^\dag )}{N}
} \!\!
\right) \!
{\displaystyle
\frac{\Phi_{00}(\widetilde{G} G^\dag )}{N}
} \!
\right\} \!
d \widetilde{G} d \widetilde{G}'  \\
\\[-12pt]
\!=\!
-
{\displaystyle
\frac{i \hbar}
{2}
}
2^{2N} \!\!
{\displaystyle \int} \!\!\!
{\displaystyle \int} \!\!
{\displaystyle
\frac{S(\!G, \widetilde{G}'\!) S(\!\widetilde{G}', \widetilde{G}\!) }
{{\Phi}_{00}(G)}
} \!
e^{-\frac{i\tau}{2}} \!\!
\left[S(\!\overline{Q}, Q\!)\right]^{\!-\frac{1}{2}} \!
{\displaystyle
\frac{\partial}{\partial t}
} \!
\left( \!
{\displaystyle
\frac{1}
{\overline{\Phi}_{00}(G)}
} 
S(\!\widetilde{G}, G\!)
e^{\frac{i\tau}{2}} \!\!
\left[S(\!\overline{Q}, Q\!)\right]^{\!-\frac{1}{2}} \!
\right) \!
d \widetilde{G} d \widetilde{G}' \\
\\[-12pt]
~+\!
{\displaystyle
\frac{i \hbar}
{2}
}
2^{2N} \!\!
{\displaystyle \int} \!\!\!
{\displaystyle \int} \!\! 
{\displaystyle
\frac{\partial}{\partial t}
} \!
\left( \!
{\displaystyle
\frac{1}
{{\Phi}_{00}(G)}
} \!
S(\!G, \widetilde{G}'\!)
e^{-\frac{i\tau}{2}} \!\!
\left[S(\!\overline{Q}, Q\!)\right]^{\!-\frac{1}{2}}  \!
\right) \!
{\displaystyle
\frac{S(\!\widetilde{G}', \widetilde{G}\!)S(\!\widetilde{G}, G\!)}
{\overline{\Phi}_{00}(G)}
} 
e^{\frac{i\tau}{2}} \!\!
\left[S(\!\overline{Q}, Q\!)\right]^{\!-\frac{1}{2}} \!
d \widetilde{G}d \widetilde{G}'  \\
\\[-12pt]
\!=\!
-
{\displaystyle
\frac{i \hbar}
{2}
} 
2^N \!\!
{\displaystyle \int} \!
{\displaystyle
\frac{1}
{{\Phi}_{00}(G)}
} 
e^{-\frac{i\tau}{2}} \!
\left[S(\!\overline{Q}, Q\!)\right]^{\!-\frac{1}{2}} \!
{\displaystyle
\frac{\partial}{\partial t}
} \!
\left( \!
{\displaystyle
\frac{S(\!G, \widetilde{G}\!) S(\!\widetilde{G}, G\!)}
{\overline{\Phi}_{00}(G)}
} 
e^{\frac{i\tau}{2}} \!
\left[S(\!\overline{Q}, Q\!)\right]^{\!-\frac{1}{2}} \!
\right) \!
d \widetilde{G} \\
\\[-12pt]
~~~\!\!\!+\!
{\displaystyle
\frac{i \hbar}
{2}
}
2^N \!\!  
{\displaystyle \int} \! 
{\displaystyle
\frac{\partial}{\partial t}
} \!
\left( \!
{\displaystyle
\frac{S(\!\widetilde{G}', G\!) S(\!G, \widetilde{G}'\!)}
{{\Phi}_{00}(G)}
} \!
e^{-\frac{i\tau}{2}} \!
\left[S(\!\overline{Q}, Q\!)\right]^{\!-\frac{1}{2}}  \!
\right) \!
{\displaystyle
\frac{1}
{\overline{\Phi}_{00}(G)}
} 
e^{\frac{i\tau}{2}} \!\!
\left[S(\!\overline{Q}, Q\!)\right]^{\!-\frac{1}{2}} \!
d \widetilde{G}'  \\
\\[-12pt]
\!=\!
-
{\displaystyle
\frac{i \! \hbar}
{2}
} \!
e^{-\frac{i\tau}{2}} \!
{\displaystyle
\frac{\left[ \! S\!(\!\overline{Q}, Q\!) \! \right]^{\!-\frac{1}{2}}}
{{\Phi}_{00}(G)}
} \!
{\displaystyle
\frac{\partial}{\partial t}
} \!\!\!
\left( \!\!
e^{\frac{i\tau}{2}} \!\!
{\displaystyle
\frac{S\!(\!G, G\!) \!\! \left[ \! S\!(\!\overline{Q}, Q\!) \! \right]^{\!-\frac{1}{2}}}
{\overline{\Phi}_{00}(G)}
} \!\!
\right)
\!\!+\!\!
{\displaystyle
\frac{i \! \hbar}
{2}
} \!
{\displaystyle
\frac{\partial}{\partial t}
} \!\!\!
\left( \!\!
e^{-\frac{i\tau}{2}} \!
{\displaystyle
\frac{\left[ \! S\!(\!\overline{Q}, Q\!) \! \right]^{\!-\frac{1}{2}}}
{{\Phi}_{00}(G)}
} \!\!
\right) \!\!
e^{\frac{i\tau}{2}} \!\!
{\displaystyle
\frac{S\!(\!G, G\!) \!\! \left[ \! S\!(\!\overline{Q}, Q\!) \! \right]^{\!-\frac{1}{2}}}
{\overline{\Phi}_{00}(G)}
}  ,
\label{Lagransian2}
\end{array}
\end{eqnarray}
using the last relation in
(\ref{Phi(00)(tildeG')1})
and
$
\frac{\dot{\Phi}_{00}(G)}{{\Phi}_{00}(G)}
\!=\!
-
i \frac{\tau}{2}
$,
which leads us to the following result:\\[-16pt]
\begin{eqnarray}
\begin{array}{l}
-
{\displaystyle
\frac{i \hbar}
{2}
} \!
\left( \!
<\!\Psi \,| \,\dot{\Psi} \!> 
-
<\!\dot{\Psi} \,| \,\Psi \!> \!
\right) \\
\\[-16pt]
=
-
{\displaystyle
\frac{i \hbar}
{2}
} 
e^{-\frac{i\tau}{2}} \!
{\displaystyle
\frac{\left[ \! S\!(\!\overline{Q}, Q\!) \! \right]^{-\frac{1}{2}}}
{{\Phi}_{00}(G)}
} 
{\displaystyle
\frac{\partial}{\partial t}
} \!
\left( \!
e^{\frac{i\tau}{2}} \!
{\Phi}_{00}(G) \!\!
\left[ \! S\!(\!\overline{Q}, Q\!) \! \right]^{\!\frac{1}{2}} \!
\right)
\!\!+\!\!
{\displaystyle
\frac{i \hbar}
{2}
} 
{\displaystyle
\frac{\partial}{\partial t}
} \!\!
\left( \!\!
e^{-\frac{i\tau}{2}} \!
{\displaystyle
\frac{\left[ \! S\!(\!\overline{Q}, Q\!) \! \right]^{\frac{1}{2}}}
{{\Phi}_{00}(G)}
} \!\!
\right) \!\!
e^{\frac{i\tau}{2}} \!
{\Phi}_{00}(G) \!\!
\left[ \! S\!(\!\overline{Q}, Q\!) \! \right]^{\frac{1}{2}} \\
\\[-18pt]
=
{\displaystyle \frac{\hbar}{2}} \dot{\tau}
-
{\displaystyle \frac{i\hbar}{4}} \!
\left( \!
{\displaystyle \dot{Q} \frac{\partial }{\partial Q}
-
\dot{\overline{Q}} \frac{\partial }{\partial \overline{Q}}} \!
\right) 
\ln S(\overline{Q}, Q)
-
{\displaystyle
\frac{i \hbar}
{2}
} 
{\displaystyle
\frac{\dot{\Phi}_{00}(G)}
{{\Phi}_{00}(G)}
}
-
{\displaystyle
\frac{i \hbar}
{2}
} 
{\displaystyle
\frac{\dot{\Phi}_{00}(G)}
{{\Phi}^2_{00}(G)}
}
{\Phi}_{00}(G)
S(\overline{Q}, Q)  \\
\\[-10pt]
=
{\displaystyle \frac{\hbar}{4}} \dot{\tau}
-
{\displaystyle \frac{i\hbar}{4}} \!
\left( \!
{\displaystyle \dot{Q} \frac{\partial }{\partial Q}
-
\dot{\overline{Q}} \frac{\partial }{\partial \overline{Q}}} \!
\right) 
\ln S(\overline{Q}, Q)
-
{\displaystyle
\frac{ \hbar}
{4}
} 
S(\overline{Q}, Q) .
\label{Lagransian3}
\end{array}
\end{eqnarray}
Using the definition
$
<\!U(G) \,| H |\,U^\dag (G) \!>
\equiv
H(G,G)
$,
the second term is also calculated as\\[-14pt]
\begin{eqnarray}
\begin{array}{l}
<\!\Psi \,| H |\,\Psi \!>
~\!\!\!\!=
2^{2N} \!\!
{\displaystyle \int} \!\!
{\displaystyle \int} \!
S(\widetilde{G}', \widetilde{G}) \! 
{\displaystyle
\frac{\overline{\Phi}_{00}(\widetilde{G}' G^\dag )}{N}
} 
H(G,G)
{\displaystyle
\frac{\Phi_{00}(\widetilde{G} G^\dag )}{N}
} 
d \widetilde{G} d \widetilde{G}'  \\
\\[-10pt]
\!\!\!\!\!\!\!\!\!\!\!\!
=
2^{2N} \!
{\displaystyle \int} \!\!
{\displaystyle \int} \!
{\displaystyle
\frac{S(\widetilde{G}', \!\widetilde{G})
S(\overline{Q}, \!\widetilde{Q}')
{\Phi}_{00}(G)
\overline{\Phi}_{00}(\widetilde{G}') }
{{\Phi}_{00}(G)}
} 
e^{-\frac{i\tau}{2}} \!
\left[S(\overline{Q}, Q)\!\right]^{-\frac{1}{2}} \\
\\[-20pt]
\times
H(G,G)
{\displaystyle
\frac{S( \overline{\widetilde{Q}}, Q )
{\Phi}_{00}(\widetilde{G})
\overline{\Phi}_{00}(G)}
{\overline{\Phi}_{00}(G)}
} 
e^{\frac{i\tau}{2}} \!
\left[S(\overline{Q}, Q)\!\right]^{-\frac{1}{2}} \!
d \widetilde{G} d \widetilde{G}' \\
\\[-10pt]
\!=\!
2^{2N} \!\!
{\displaystyle \int} \!\!\!
{\displaystyle \int} \!\!
{\displaystyle
\frac{S(\!G, \widetilde{G}'\!) S(\!\widetilde{G}', \widetilde{G}\!) }
{{\Phi}_{00}(G)}
} \!
e^{-\frac{i\tau}{2}} \!\!
\left[S(\!\overline{Q}, Q\!)\right]^{\!-\frac{1}{2}} \!
H(G,G) ~\!
{\displaystyle
\frac{S(\!\widetilde{G}, G\!)}
{\overline{\Phi}_{00}(G)}
} 
e^{\frac{i\tau}{2}} \!\!
\left[S(\!\overline{Q}, Q\!)\right]^{\!-\frac{1}{2}} \!
d \widetilde{G} d \widetilde{G}' \\
\\[-10pt]
\!=\!
2^N \!\! 
{\displaystyle \int}
{\displaystyle
\frac{S(\!G, \widetilde{G}\!) }
{{\Phi}_{00}(G)}
} 
e^{-\frac{i\tau}{2}} \!
\left[S(\!\overline{Q}, Q\!)\right]^{\!-\frac{1}{2}} \!
H(G,G) ~\!
{\displaystyle
\frac{S(\!\widetilde{G}, G\!)}
{\overline{\Phi}_{00}(G)}
} 
e^{\frac{i\tau}{2}} \!
\left[S(\!\overline{Q}, Q\!)\right]^{\!-\frac{1}{2}} \!
d \widetilde{G} \\
\\[-12pt]
\!=\!
e^{-\frac{i\tau}{2}} \!
\left[S(\!\overline{Q}, Q\!)\right]^{\!-\frac{1}{2}} \!
{\displaystyle
\frac{H(G,G)}{\left|\Phi_{00}(G)\right|^2}
}
e^{\frac{i\tau}{2}} \!
\left[S(\!\overline{Q}, Q\!)\right]^{\!-\frac{1}{2}}
\!=\!
{\displaystyle
\frac{H(\overline{Q}, Q)}{S(\overline{Q}, Q)}
} .
\label{Lagransian4}
\end{array}
\end{eqnarray}\\[-8pt]
Using
(\ref{Lagransian})
with the aid of
(\ref{Lagransian3})
and
(\ref{Lagransian4}),
we finally obtain
a classical TD $SO(2N \!+\! 1)$ Lagrangian
given as\\[-18pt]
\begin{eqnarray}
\begin{array}{l}
{\cal L}
=
{\displaystyle
\frac{H(\overline{Q}, Q)}{S(\overline{Q}, Q)}
}
-
{\displaystyle \frac{i\hbar}{4}} \!
\left( \!
{\displaystyle \dot{Q} \frac{\partial }{\partial Q}
-
\dot{\overline{Q}} \frac{\partial }{\partial \overline{Q}}} \!
\right) \!
\ln S(\overline{Q}, Q)
-
{\displaystyle
\frac{ \hbar}
{4}
} 
S(\overline{Q}, Q)
+
{\displaystyle \frac{\hbar}{4}} \dot{\tau} .
\end{array}
\label{L(Q'Q)}
\end{eqnarray}\\[-10pt]
The Euler-Lagrange equation of motion for the
$\!\frac{SO(2N \!+\! 1)}{U(N \!+\! 1)}\!$
coset variable $\overline{Q}_{\!\alpha \beta}\!$
is calculated to be\\[-16pt]
\begin{eqnarray}
\begin{array}{l}
{\displaystyle \frac{d}{dt}} \!
\left( \!
{\displaystyle 
\frac{\partial \cal L}{\partial \dot{\overline{Q}}_{pq}} \!
}
\right)
\!-\!
{\displaystyle 
\frac{\partial \cal L}{\partial \overline{Q}_{pq}}
} \\
\\[-10pt]
=
{\displaystyle \frac{i\hbar}{4}} 
{\displaystyle \frac{\partial}{\partial t}} \!
\left\{ \!
{\displaystyle 
\frac{\partial}{\partial \overline{Q}_{pq}}
} \!
\ln S(\overline{Q}, Q) \!
\right\}
\!\!-\!\!
{\displaystyle 
\frac{\partial}{\partial \overline{Q}_{pq}}
} \!\!
\left[ \!
{\displaystyle
\frac{H(\overline{Q}, Q)}{S(\overline{Q}, Q)}
} \!
\right]
\!\!+\!\!
{\displaystyle \frac{i\hbar}{4}} \!
\left( \!\!
{\displaystyle \dot{Q}_{rs} \frac{\partial }{\partial Q_{rs}}
-
\dot{\overline{Q}}_{rs} \frac{\partial }{\partial \overline{Q}_{rs}}} \!\!
\right) \!\!
{\displaystyle 
\frac{\partial}{\partial \overline{Q}_{pq}}
} \!
\ln S(\overline{Q}, Q) \\
\\[-10pt]
~\!\!\!+
{\displaystyle \frac{\hbar}{4}} 
{\displaystyle
\frac{\partial S(\overline{Q}, Q)}{\partial \overline{Q}_{pq}}
} \\
\\[-10pt]
=
-
{\displaystyle 
\frac{\partial}{\partial \overline{Q}_{pq}}
} \!\!
\left[ \!
{\displaystyle
\frac{H(\overline{Q}, Q)}{S(\overline{Q}, Q)}
} \!
\right]
+
{\displaystyle \frac{i\hbar}{2}} 
\dot{Q}_{rs}
{\displaystyle 
\frac{\partial^2}{\partial {Q}_{rs}\partial \overline{Q}_{pq}}
} \!
\ln S(\overline{Q}, Q)
+
{\displaystyle \frac{\hbar}{4}} 
{\displaystyle
\frac{\partial S(\overline{Q}, Q)}{\partial \overline{Q}_{pq}}
} \\
\\[-10pt] 
=
0 ,
\end{array}
\label{Euler-Lagrange}
\end{eqnarray}\\[-6pt]
and its complex conjugation.
Equation
(\ref{Euler-Lagrange})
leads to
the different form of the $SO(2N \!\!+\!\! 1)$ TDHB equation
from the one of the TDHB equation given in Ref.
\cite{Fuku.Nishi.84}.
$\!$This form of equation
also describes collective excitations in even and odd fermion-number systems,
respectively.


\newpage


\def\thesection{\arabic{section}}
\setcounter{equation}{0}
\renewcommand{\theequation}{\arabic{section}.
\arabic{equation}}
\section{RPA vacuum  and symplectic two-form $\omega$}


~~
We are now in a stage to derive an equation for 
the $SO(2N \!+\! 1)$ random phase approximation (RPA)
which describes collective excitations in even and odd Fermion systems,
respectively.
Let $\cal G, \cal G'\!$ and $\!{\cal G}_0$  be the $SO(2N \!\!+\!\! 2)$  matrices
corresponding to $G, G'\!$ and $\!G_0$, respectively.
The $\cal G$ is given by
(\ref{calG}).
Following Fukutome
\cite{pricommuF},
we consider the fluctuation $\widetilde{G}$ around
the stationary ground state $|G_0 \rangle$.
It can be regarded as
the matrix-valued generator coordinate  
in the $\cal G$ quasi-particle frame defined by
the following relation  
$\widetilde{\cal G} \!=\! {\cal G}_0^{\dagger}{\cal G}$:

 As shown previously, a state vector in the $SO(2N \!+\! 1)$ spinor space is
 in the form of\\[-20pt] 
\begin{eqnarray}
\begin{array}{c}
\Phi_{0f}(G_0^\dag G) 
= 
\Phi_{00}(G_0^\dag G) 
\chi_{\!f} [\overline{\widetilde{\cal Q}} (G_0^\dag G) ],~
\Phi_{00}(\widetilde{\cal G})
\!=\!
\Phi_{00}(\widetilde{G}) .
\end{array}
\label{Phi(0f)(gg0dag)}
\end{eqnarray}\\[-18pt]
Using the same representations as those of
(\ref{AandB})
for the relation
$\widetilde{\cal G} \!=\! {\cal G}_0^{\dagger}{\cal G}$,
we have\\[-16pt]
\begin{eqnarray} 
\begin{array}{rl}
\widetilde{\cal G}
\!=\! 
\left[ \!\!
\begin{array}{cc}
\widetilde{\cal A} & \overline{\widetilde{\cal B}} \\
\widetilde{\cal B} & \overline{\widetilde{\cal A}}
\end{array} \!\!
\right]
\!\!\!  
&
\!=\! 
{\cal G}^\dag_0 {\cal G}    
\!=\!
\left[ \!\!
\begin{array}{cc}
{\cal A}^\dag_0&{\cal B}^\dag_0\\
\\[-10pt]
{\cal B}^{\mbox{\scriptsize T}}_0&{\cal A}^{\mbox{\scriptsize T}}_0
\end{array} \!\!
\right] \!
\left[ \!
\begin{array}{cc}
{\cal A}&\overline{\cal B}\\
\\[-10pt]
{\cal B}&\overline{\cal A}
\end{array} \!
\right] \!
\!=\!
\left[ \!
\begin{array}{cc}
{\cal A}^\dag_0 {\cal A} \!+\! {\cal B}^\dag_0 {\cal B}&
{\cal A}^\dag_0 \overline{\cal B} \!+\! {\cal B}^\dag_0 \overline{\cal A}\\
\\[-10pt]
{\cal B}^{\mbox{\scriptsize T}}_0 {\cal A}
\!+\!
{\cal A}^{\mbox{\scriptsize T}}_0 {\cal B}&
{\cal B}^{\mbox{\scriptsize T}}_0 \overline{\cal B}
\!+\!
{\cal A}^{\mbox{\scriptsize T}}_0 \overline{\cal A}
\end{array} \!
\right] .
\end{array}
\label{coordinatecalG0}
\end{eqnarray}\\[-10pt]
Then, with the aid of the last relation of
(\ref{noneuclidiantransformation0}),
a variable
$
\widetilde{\cal Q}
\!\equiv\!
\widetilde{\cal B} \widetilde{\cal A}^{-1}
$
is written as\\[-18pt]
\begin{eqnarray} 
\begin{array}{ll}
\widetilde{\cal Q}
&\!\!\!=\!
\widetilde{\cal B} \widetilde{\cal A}^{-1}
\!=\!
\left( 
{\cal B}^{\mbox{\scriptsize T}}_0 {\cal A}
\!+\!
{\cal A}^{\mbox{\scriptsize T}}_0 {\cal B} 
\right) \!
( 
{\cal A}^\dag_0 {\cal A} \!+\! {\cal B}^\dag_0 {\cal B}
)^{-1}
\!=\!
\left( 
{\cal B}^{\mbox{\scriptsize T}}_0 \!+\! {\cal A}^{\mbox{\scriptsize T}}_0 {\cal Q} 
\right) \!
(
{\cal A}^\dag_0 \!+\! {\cal B}^\dag_0 {\cal Q}
)^{-1} \\
\\[-10pt]
&\!\!\!=\!
(\overline{\cal A}_0)^{-1} \!\!
\left( 
\overline{\cal A}_0 {\cal B}^{\mbox{\scriptsize T}}_0
\!+\!
\overline{\cal A}_0 {\cal A}^{\mbox{\scriptsize T}}_0 {\cal Q} 
\right) \!\!
\left( 1_{\!N\!+\!1} \!\!-\!\! \overline{\cal Q}_0 {\cal Q} \right)^{\!-1} \!\!
({\cal A}^\dag_0)^{-1} \\
\\[-10pt] 
&\!\!\!=\!
(\overline{\cal A}_0)^{-1} \!\!
\left[
\left\{ \!
{\cal Q}
\!+\!
\overline{\cal A}_0 {\cal B}^{\mbox{\scriptsize T}}_0
\!-\!
{\cal B}_0 {\cal B}^\dag_0 
{\cal Q}
\right\} \!\!
\left( 1_{\!N\!+\!1} \!\!-\!\! \overline{\cal Q}_0 {\cal Q} \right)^{\!-1} 
\right] \!\!
({\cal A}^\dag_0)^{-1} \\
\\[-10pt]
&\!\!\!=\!
(\overline{\cal A}_0)^{-1} \!
\left[ 
-{\cal Q}_0
\left( 1_{\!N\!+\!1} \!-\! \overline{\cal Q}_0 {\cal Q}_0 \right)^{-1}
\!+\!
{\cal Q}
\left( 1_{\!N\!+\!1} \!-\! \overline{\cal Q}_0 {\cal Q} \right)^{-1} 
\right] \!
({\cal A}^\dag_0)^{-1} ,
\end{array}
\label{coordinatecalQ}
\end{eqnarray}\\[-12pt]
where we have used
${\cal Q} \!\equiv\! {\cal B}{\cal A}^{-1}$.
To reach the last line of
(\ref{coordinatecalQ}),
further we have used the relations
$
\overline{\cal A}_0 {\cal A}^{\mbox{\scriptsize T}}_0
\!+\!
{\cal B}_0 {\cal B}^\dag_0
\!=\!
1_{\!N\!+\!1}
$,
$
\overline{\cal A}_0 {\cal B}^{\mbox{\scriptsize T}}_0
\!=\!
-{\cal Q}_0
\left( \! 1_{\!N\!+\!1} \!-\! \overline{\cal Q}_0 {\cal Q}_0 \! \right)^{-1}
\!$
and
$
{\cal B}_0 {\cal B}^\dag_0
\!=\!
-{\cal Q}_0 \!
\left( \! 1_{\!N\!+\!1} \!-\! \overline{\cal Q}_0 {\cal Q}_0 \! \right)^{-1} \!
\overline{\cal Q}_0 
$,
in which the last two relations can be derived from
(\ref{matA0andB0}).
At first glance,
the transformation rule
(\ref{coordinatecalQ})
seems to be very different from the one
(\ref{noneuclidiantransformation})
given in the previous  section.
This is because we here choose
the matrix-valued generator coordinate  
in the $\cal G$ quasi-particle frame defined by the relation
$\widetilde{\cal G} \!=\! {\cal G}_0^{\dagger}{\cal G}$
though before we adopted
the matrix-valued generator coordinate  
in the $\cal G$ quasi-particle frame defined by another relation
$\widetilde{\cal G} \!=\! {\cal G}^{\dagger}{\cal G}'$.

Hereafter we restrict the $G_0$ to
a $g_0$, the $SO(2N)$ HB case.
Then, from
(\ref{matA0andB0}),
we have\\[-20pt]
\begin{eqnarray} 
\begin{array}{c}
\overline{\widetilde{\cal Q}}_{ij}
\!=\!
\overline{\widetilde{q}}_{ij}
\!=\!
\left[
a_0^{-1} \!
\left\{
\overline{q} \!
\left( 1 \!-\! q_0 \overline{q} \right)^{-1}
\!-\!
\overline{q}_0 \!
\left( 1 \!-\! q_0 \overline{q}_0 \right)^{\!-1} \!
\right\} \!
a^{\mbox{\scriptsize T}-1}_0
\right]_{ij} , 
\end{array}
\label{coordinatecalqHB}
\end{eqnarray}\\[-20pt]
denoting $\widetilde{q}$ simply as $q$.
In that case,
the Dyson rep
(\ref{SO2Nplus1Lieopa2})
leads to the operators
$\mbox{\boldmath $\widetilde{e}^i_{~j }$}\!$ etc. for
even number systems
and the commutation relation
$[\mbox{\boldmath $\widetilde{e}_{ij }$}, \mbox{\boldmath $\widetilde{e}^{lk}$}]$
in the following forms:\\[-22pt]
\beqa
\left.
\BA{cc}
&
{\displaystyle
\mbox{\boldmath $\widetilde{e}^i_{~j }$}
\!+\!
\frac{1}{2} \delta_{ij}
\!=\!
\overline{q}_{ik} \frac{\partial }{\partial \overline{q}_{jk }}
},~~
\mbox{\boldmath $\widetilde{e}_{ij }$}
\!=\!
-{\displaystyle\frac{\partial }{\partial \overline{q}_{ij }}} ,~~
\mbox{\boldmath $\widetilde{e}^{ij }$}
\!\simeq\!
\overline{q}_{ij },~\\
\\[-10pt]
&
[\mbox{\boldmath $\widetilde{e}_{ij }$}, \mbox{\boldmath $\widetilde{e}^{lk}$}]
\!\simeq\!
\delta_{ik} \delta_{jl} - \delta_{il}\delta_{jk} .
\EA \!\!\!
\right\}
\label{SO2NHBLieopa}
\eeqa\\[-26pt]

Let us introduce new boson annihilation-creation operators
$\frac{\partial }{\partial \omega_A}$
and
$\omega_A$.
Then, the old ones
$\frac{\partial }{\partial \overline{q}_{ij }}$
and
$\overline{q}_{ij }$
are expressed in the linear combination of the new ones
as follows:\\[-22pt]
\beqa
\left\{ \!\!\!
\BA{lr}
&\!\!\!\!
{\displaystyle
\overline{q}_{ij }
\!=\!
\overline{u}_{ij,A } \omega_A
+
\overline{v}_{ij,A } \frac{\partial }{\partial \omega_A}
},~ 
(i>j), \\
\\[-16pt]
&\!\!\!\!
{\displaystyle
\frac{\partial }{\partial \overline{q}_{ij }}
\!=\!
u_{ij,A } \frac{\partial }{\partial \omega_A}
+
v_{ij,A } \omega_A 
},~
(i>j), 
\EA 
\right.
\label{HBqLieopa1}
\eeqa\\[-14pt]
where the coefficients $u$ and $v$
satisfy the following orthogonal conditions:\\[-16pt]

\beqa
\left\{ \!\!\!
\BA{lr}
&\!\!\!\!
u_{ij,A } \overline{u}_{kl,A } 
-
v_{ij,A } \overline{v}_{kl,A }
\!=\!
\delta_{ij, kl} ,~
\left(
uu^\dag \!-\! vv^\dag \!=\! 1_{\!N}
\right) , \\
\\[-6pt]
&\!\!\!\!
v_{ij,A } u_{kl,A } 
-
v_{kj,A } u_{kl,A }
\!=\!
0 ,~~~~~
\left(
vu^{\mbox{\scriptsize T}} \!-\! uv^{\mbox{\scriptsize T}}
\!=\!
0
\right) .
\EA \!\!\!\!
\right.
\label{HBuvortho1}
\eeqa\\[-12pt]
In
(\ref{HBqLieopa1})
and
(\ref{HBuvortho1}),
we have used the summation convention over repeated index $A$.
The inverse transformation to
(\ref{HBqLieopa1})
is given as follows:\\[-22pt]
\beqa
\left\{ \!\!\!
\BA{lr}
&\!\!\!\!
\omega_A
\!=\!
\sum_{i>j} \!
\left( \!
u_{ij,A } \overline{q}_{ij }
+
\overline{v}_{ij,A } {\displaystyle\frac{\partial }{\partial \overline{q}_{ij }}} \!
\right) , \\
\\[-12pt]
&\!\!\!\!
{\displaystyle
\frac{\partial }{\partial \omega_A}
}
\!=\!
\sum_{i>j} \!
\left( \!
\overline{u}_{ij,A } {\displaystyle \frac{\partial }{\partial \overline{q}_{ij }}}
+
v_{ij,A } \overline{q}_{ij } \!
\right) ,
\EA 
\right.
\label{HBqLieopa2}
\eeqa\\[-14pt]
where the coefficients $u$ and $v$
satisfy the other type of the orthogonal conditions:\\[-18pt]
\beqa
\left\{ \!\!\!
\BA{ll}
&\!\!\!\!
\sum_{i>j}
\left( \!
\overline{u}_{ij,A } u_{ij,B }
-
v_{ij,A } \overline{v}_{ij,B }
\right)
\!=\!
\delta_{AB} ,~
\left(
u^\dag u \!-\! v^{\mbox{\scriptsize T}} \overline{v} \!=\! 1_{\!N}
\right) , \\
\\[-6pt]
&\!\!\!\!
\sum_{i>j}
\left( \!
\overline{u}_{ij,A } \overline{v}_{ij,B }
-
u_{ij,B } \overline{v}_{ij,A }
\right)
\!=\!
0 ,~~~~
\left(
u^{\mbox{\scriptsize T}} \overline{v} \!-\! v^\dag u \!=\! 0
\right) . 
\EA \!\!\!\!\!\!\!\!\!\!\!\!\!\!\!\!\!\!
\right.
\label{HBuvortho2}
\eeqa\\[-26pt]

We here put $f \!=\! 0$ in the function $\chi_{\!f}$ given in
(\ref{Phi(0f)(gg0dag)}).
This means that
we treat only the ground state
with no bosons.
Then, the function $\chi_0$ should satisfy the following condition:\\[-20pt]
\beqa
\BA{c}
{\displaystyle
\frac{\partial \chi_0}{\partial \omega_A}
}
=
\sum_{i>j} \!
\left( \!
\overline{u}_{ij,A } {\displaystyle \frac{\partial }{\partial \overline{q}_{ij }}}
+
v_{ij,A } \overline{q}_{ij } \!
\right) \!
\chi_0
=
0 ,
\EA
\label{RPAvac}
\eeqa\\[-14pt]
where we have used the second equation of
(\ref{HBqLieopa2}).
From
(\ref{RPAvac}),
we can get the differential equation for $\chi_0$
with respect to the variable $\overline{q}_{ij }$
as\\[-20pt]
\beqa
\left\{ \!\!\!\!\!\!
\BA{ll}
&
{\displaystyle
\frac{\partial \chi_0}{\partial \overline{q}_{ij }}
}
\!=\!
-
\sum_{k>l} 
R_{ij, kl } \overline{q}_{kl} \chi_0 , \\
\\[-8pt]
&
R_{ij, kl }
\!\equiv\!
\left[
(u^\dag)^{-1} v^{\mbox{\scriptsize T} } 
\right]_{ij, kl }
\!=\!
\left[
v (\overline{u})^{-1}
\right]_{ij, kl }, ~
R_{ij, kl }
\!=\!
R_{kl,  ij} .
\EA \!\!\!
\right.
\label{chiqdiff}
\eeqa\\[-10pt]
the solution for which
is easily obtained as\\[-22pt]
\beqa
\BA{c}
\chi_0
=
\exp
\left[
- {\displaystyle \frac{1 }{2}}
\sum_{i > j, ~k > l }
R_{ij, kl }\overline{q}_{ij } \overline{q}_{kl}
\right] ,
\EA
\label{RPAvacuum}
\eeqa\\[-16pt]
which leads to the RPA vacuum 
$
\Phi_0 
\!=\! 
\Phi_{00}(G^\dag_0 G) 
\chi_0 
$.

On the other hand,
by using the boson creation operator $\omega_A$
(\ref{HBqLieopa2}),
the RPA excited states
with one boson and two bosons can be realized,
respectively,
in the following forms:

One boson excited state:\\[-22pt]
\begin{eqnarray}
\begin{array}{ll}
\omega_A \chi_0
&\!\!\!\!=\!
\sum_{i>j} \!
\left( \!
u_{ij,A } \overline{q}_{ij }
+
\overline{v}_{ij,A } {\displaystyle\frac{\partial }{\partial \overline{q}_{ij }}} \!
\right) \!
\chi_0 \\
\\[-12pt]
&\!\!\!\!=\!
\sum_{i>j} \!
\left( 
u_{ij,A }
\!-\!
\sum_{k>l} \overline{v}_{kl,A } R_{kl, ij } 
\right) \!
\overline{q}_{ij }
\chi_0 
\!=\!
\sum_{i>j} \!
\left[
u^{\mbox{\scriptsize T} }
\!-\!
 v^\dag v (\overline{u})^{-1}
\right]_{ij,A } \!
\overline{q}_{ij }
\chi_0 \\
\\[-8pt]
&\!\!\!\!=\!
\left[(\overline{u})^{-1}~\!\overline{q}\right]_A \!
\chi_0
\end{array}
\label{RPAexcited1}
\end{eqnarray}\\[-12pt]
where we have used the definition of $R_{kl, ij }$
given by the second equation of
(\ref{chiqdiff}).

Two bosons excited state:\\[-24pt]
\begin{eqnarray}
\begin{array}{l}
\omega_A \omega_B \chi_0
\!=\!
\left\{
\left[(\overline{u})^{-1}~\!\overline{q}\right]_A
\left[(\overline{u})^{-1}~\!\overline{q}\right]_B
\!+\!
\left[(\overline{u})^{-1}~\!\overline{v}\right]_{AB}
\right\} \!
\chi_0 .
\end{array}
\label{RPAexcited2}
\end{eqnarray}\\[-34pt]

Finally, we point out the non existence of the higher RPA vacuum.
Suppose the function $\chi$ corresponds to
the higher RPA vacuum.
Then, the $\chi$ contains no excited bosons.
This is shown as follows:
The function $\chi$ should satisfy the condition\\[-20pt] 
\begin{eqnarray}
\begin{array}{l}
\sum_{i>j} 
\left( 
\overline{u}_{ij,A } \mbox{\boldmath $\widetilde{e}_{ij }$}
-
v_{ij,A } \mbox{\boldmath $\widetilde{e}^{ij }$} 
\right) \!
\chi
=
0 ~
\longrightarrow
\mbox{\boldmath $\widetilde{e}_{ij }$} \chi
\!=\!
R_{ij, kl }
\mbox{\boldmath $\widetilde{e}^{kl }$} \chi .
\end{array}
\label{higherRPAexcited1}
\end{eqnarray}\\[-18pt]

Using the Dyson rep
(\ref{SO2Nplus1Lieopa2}),
this condition
(\ref{higherRPAexcited1})
is transformed to\\[-18pt]
\begin{eqnarray}
\begin{array}{l}
\left\{ \!
{\displaystyle\frac{\partial }{\partial \overline{q}_{ij }}}
\!+\!
R_{ij, kl }
\left( \!
\overline{q}_{km} \overline{q}_{nl }
{\displaystyle\frac{\partial }{\partial \overline{q}_{mn }}}
\!+\!
\overline{q}_{kl } \!
\right) \!
\right\} \!
\chi
\!=\!
0 ~
\longrightarrow
(1 - R~\overline{q}^2)
{\displaystyle\frac{\partial }{\partial \overline{q}_{ij }}} \chi
\!=\!
- R
\overline{q} \chi , 
\end{array}
\label{higherRPAexcited2}
\end{eqnarray}\\[-14pt]
from which we have 
\beq
{\displaystyle\frac{\partial }{\partial \overline{q}_{ij }}} \!
\ln \! \chi
\!=\!
-
\left[
(1 \!\!-\!\! R \overline{q}^2)^{\!-1} \! R \overline{q}~\!
\right]_{ij } .
\label{fij} 
\eeq\\[-10pt]
Further
the second differential for $\ln \! \chi$ with respect to $\overline{q}$
is computed as\\[-18pt]
\begin{eqnarray}
\!\!\!\!\!\!\!\!
\begin{array}{ll}
{\displaystyle\frac{\partial^2 }{\partial \overline{q}_{kl } \partial \overline{q}_{ij }}} \!
\ln \! \chi
\!\!\!\!\!&~\!\!
=
{\displaystyle\frac{\partial f_{ij}}{\partial \overline{q}_{kl }}} \\
\\[-16pt]
&~\!\!
=
-
\left[
(1 \!\!-\!\! R~\overline{q}^2)^{-1} \! R~\! 
\right]_{ij,~kl }
\!-\!
2
\left[
(1 \!\!-\!\! R~\overline{q}^2)^{-1} \! R~\! 
\right]_{ij,~lm} \!
\overline{q}_{mn} \overline{q}_{nl } \!
\left[
(1 \!\!-\!\! R~\overline{q}^2)^{-1} \! R~\overline{q}
\right]_{ln} \\
\\[-16pt]
&~\!\!
\!\neq\!
{\displaystyle\frac{\partial^2 }{\partial \overline{q}_{ij } \partial \overline{q}_{kl } }} \!
\ln \! \chi ,
\end{array}
\label{higherRPAnonintegral}
\end{eqnarray}\\[-12pt]
which shows $\ln \! \chi$ not to be integrable.
An integrability condition is intimately related to the Pfaff's problem
\cite{Forsyth.1885}
completely solved by Cartan
\cite{Cartan.45}. 
$\!$Thus
the non existence of the higher RPA vacuum is proved. 
$\!\!$A curvature
$C(\!=\! d \Omega \!-\! \Omega \!\wedge\! \Omega)$ 
to become zero is nothing but the vanishing of the curvature $C\!$ of connection.
$\!$A one-form $\Omega$
is linearly composed of some infinitesimal generators
\cite{Nishi.Provi.Komatsu.09}.
The one boson excited state on the the function $\chi$ is constructed
as\\[-16pt]
\begin{eqnarray}
\begin{array}{ll}
\omega_A \chi_0
&\!\!\!\!=\!
\sum_{i>j} \!
\left( \!
u_{ij,A } \mbox{\boldmath $\widetilde{e}_{ij }$}
-
\overline{v}_{ij,A } \mbox{\boldmath $\widetilde{e}^{ij }$} \!
\right) \!
\chi_0 \\
\\[-10pt]
&\!\!\!\!=\!
\sum_{i>j} \!
\left\{ 
u_{ij,A }
\left( \!
\overline{q}_{ij } 
\!+\!
\overline{q}_{ik} \overline{q}_{lj}
{\displaystyle\frac{\partial }{\partial \overline{q}_{kl }}} \!
\right) 
\!+\!
\overline{v}_{ij,A }
{\displaystyle\frac{\partial }{\partial \overline{q}_{ij}}} \!
\right\} \!
\chi \\
\\[-10pt]
&\!\!\!\!=\!
\sum_{i>j} \!
\left\{ 
u_{ij,A }
\left( 
\overline{q}_{ij } 
\!-\!
\overline{q}_{ik} \overline{q}_{lj}
\overline{f}_{\!kl } 
\right) 
\!-\!
\overline{v}_{ij,A }
f_{\!ij} \!
\right\} \!
\chi .
\end{array}
\label{higherRPAoneBoson}
\end{eqnarray}\\[-22pt]

$\!\!\!\!$From the group theoretical viewpoint$\!$
\cite{Nishi.Provi.Komatsu.09},
we show the existence of homogeneous symplectic
two-form
$\omega$.
We introduce a hermitian and traceless HB density-matrix $w$
on the $\frac{SO(2N+2)}{U(N+1)}$ \\[-16pt]
\begin{eqnarray}
\!\!\!\!\!\!\!\!\!\!
\begin{array}{c}
w
\!=\!
\overline{\cal G}
\epsilon
{\cal G}^{\mbox{\scriptsize T}}
\!\!=\!
\overline{\cal G} \!\!
\left[ \!\!\!
\begin{array}{cc}
-1_{\!N\!+\!1} &\!\!\!\! 0 \!\!\!\\
\\[-6pt]
0 &\!\!\!\! 1_{\!N\!+\!1} \!\!\!
\end{array} 
\right] \!\!
{\cal G}^{\mbox{\scriptsize T}}
\!\!=\!\!
\left[ \!\!\!
\begin{array}{cc}
2 {\cal B}{\cal B}^\dag \!\!-\!\! 1_{\!N\!+\!1} &\!\!\!\!\!\!
2 {\cal B}{\cal A}^\dag \\
\\[-6pt]
-2 \overline{\cal B} {\cal A}^{\mbox{\scriptsize T}}&\!\!\!\!\!\!
-2 \overline{\cal B} {\cal B}^{\mbox{\scriptsize T}}
\!\!+\!\! 1_{\!N\!+\!1} \!\!\!
\end{array} 
\right] \!
\!=\!
2 {\cal W} \!-\! 1_{\!2N\!+\!2}, ~
w^2
\!\!=\!\!
1_{\!2N\!+\!2} ,
\end{array} \!\!
\label{densitywg}
\end{eqnarray}\\[-10pt]
with $w dw \!+\! dw w \!=\! 0$.
The action of $SO(2N\!+\!2)$ on $w$ is made as
$w \!\rightarrow\! \overline{\cal G}w{\cal G}^{\mbox{\scriptsize T}}$. 
Taking a new $(N\!+\!1) \!\times\! (2N\!+\!2)$ matrix $h$ as
$
h^\dag
\!=\!
\left(1_{\!N\!+\!1} \!+\! {\cal Q} {\cal Q}^\dag \right)^{\!-\frac{1}{2}} \!
\left[ 1_{\!N\!+\!1}, -{\cal Q} \right]
$
with
$
h^\dag h
\!=\!
1_{\!2N\!+\!2}
$,
we have a very simple expression for $w$ as\\[-18pt]
\beqa
w
=
1_{\!2N\!+\!2}
-
2 h h^\dag ,~
\mbox{Tr}w
=
0 ,
\label{densitywg2}
\eeqa\\[-18pt]
Following Rajeev, Toprak and Tugurt
\cite{Rajeev.94,RajeevTurgut.98,ToprakTugurt.02,ToprakTugurt2.02},
the \underline{invariant} $\omega$
and $d \omega$
are given as\\[-10pt]
\beq
\omega
\!\!=\!\!
-
\frac{i}{8} \!
\mbox{Tr}\!\!
\left\{ \! w \! \left(\!dw\!\right)^{\!2} \! \right\}\!
(\mbox{\underline{invariant} under}~w
\!\!\rightarrow\!\!
\overline{\cal G}w{\cal G}^{\mbox{\scriptsize T}}\!),~
d \omega
\!\!=\!\!
-
\frac{i}{8} \!
\mbox{Tr}\!\!
\left\{ \! \left(\!dw\!\right)^{\!3} \! \right\}
\!\!=\!\!
-
\frac{i}{8} \!
\mbox{Tr}\!\!
\left\{ \! \left(\!dw\!\right)^{\!3} \!\! w^{\!2} \! \right\}
\!\!=\!\!
-
d\omega ,
\label{twoform}
\eeq\\[-14pt]
and then,
$
d\omega
\!=\!
0~(\mbox{closed form}).
$
If we introduce hermitian matrices
$
\Psi 
\!\!=\!\!
\left[ \!\!\!
\begin{array}{cc}
0 & \!\!\!\! \psi \!\!\! \\
\psi^\dag & \!\!\!\! 0 \!\!\!
\end{array} 
\right]
\!$
and
$
\Phi
\!\!=\!\!
\left[ \!\!
\begin{array}{cc}
0 & \!\!\!\! \phi \!\!\! \\
\phi^\dag & \!\!\!\! 0 \!\!\!
\end{array}
\right] \! ,
$
tangent vectors satisfying
$\{\epsilon, \Psi\}\!=\!\{\epsilon, \Phi\}\!=\!0$,
thus, we have the Rajeev's quadratic form:\\[-6pt]
\beq
\omega(\Psi,~\Phi)
\!=\!
-
\frac{i}{8}
\mbox{Tr}
\left\{
\left[ \!\!
\begin{array}{cc}
1_N & \!\!\!\! 0 \\
0 & \!\!\!\! -1_N
\end{array} \!\!
\right] \!
[\Psi,~\Phi] 
\right\}
\!=\!
\frac{i}{4}
\mbox{Tr}
\left\{ \psi^\dag \varphi - \varphi^\dag \psi \right\} ,
\label{symplectictwoform}
\eeq
which is a symplectic and nondegenerate form.
The Grassmannian $w$ is a symplectic manifold with
the two-form $\omega$.
A symplectic vector field $V_{\!\psi}(w)$
satisfying
$wV_{\!\psi}(w) \!+\! V_{\!\psi}w \!\!=\!\! 0$
is found to be
$V_{\!\psi}(w) \!\!=\!\! - i [\psi, w]$.
This is the action of $V_{\!\psi}(w)$
on the $w$ for a Lie algebra element $\psi(w)$.
The symplectic structure can be considered in terms of
the Poisson algebra of these functional matrices.
Standing on the above observation,
it makes possible to construct the geometric quantization on
a finite-dimensional Grassmannian
as has been discussed in Ref.
\cite{Rajeev.94}.


\newpage

\setcounter{equation}{0}
\renewcommand{\theequation}{\arabic{section}.\arabic{equation}}

\section{Discussions and further perspective}

\vspace{-0.2cm} 

~~~~A different form of the $SO(2N \!\!+\!\! 1)$ TDHB theory
from the previous one of Ref.
\cite{Fuku.Nishi.84}
has been made, basing on the fact that
the fermion annihilation-creation and pair operators form
the Lie algebra of $SO(2N \!\!+\!\! 1)$ group.
The $SO(2N \!+\! 1)$ group was found by a group
extension of the $SO(2N)$ Bogoliubov transformation for the fermion
to a new canonical transformation group.
Embedding the $SO(2N \!\!+\!\! 1)$ group into the $SO(2N \!\!+\!\! 2)$ group and
using the $\frac{SO(2N \!+\! 2)}{U(N \!+\! 1)}$ coset variables,
we have developed the extended TDHB theory in which
paired and unpaired modes are treated in an equal manner.
Such the TDHB theory applicable to both even and odd fermion-number systems
is also the $\!$TDSCF theory with the same level of the mean field approximation
as the usual $\!$TDHB theory for even fermion-number systems
\cite{RS.80,BR.86}.

Adopting a slight different way from Fukutome's
\cite{Fuku.78},
we here have proposed the non-Euclidian transformation,
bringing the projected $SO(2N \!\!+\!\! 1)$ Tamm-Dancoff equation
of Ref.
\cite{Fuku.78}
and derived the classical TD $SO(2N \!\!+\!\! 1)$ Lagrangian
which,
through the Euler-Lagrange equation
of motion for $\frac{SO(2N \!\!+\!\! 2)}{U(N \!\!+\!\! 1)}$ coset variables,
leads to
a different form of the extended TDHB equation
from that of Ref.
\cite{Fuku.Nishi.84}.
The RPA, starting with HB approximation
\cite{Bogo.58,Bogo.Tol.Sir.58}
for a ground state,
HB RPA,
has been applied to superconducting fermion systems.
The HB RPA, however, is applicable only to even fermion-number systems
because the HB WF contains only components
with even fermion numbers
and describes only Bose type excitations
which are ascribed to creation and annihilation of quasi-particle pairs.
The present RPA is derived using the Dyson rep
\cite{Fuku.77}
for paired and unpaired operators.
In the $SO(2N)$ HB case,
the RPA vacuum has been obtained.
One boson and two boson excited states have also been realized.
We, however, have stressed the non existence of the higher RPA vacuum
because the $SO(2N \!\!+\!\! 1)$
spinor function is not integrable
and existence of the symplectic two-form
$\omega$.

$\!\!$Being described in the previous section,
the symplectic structure is considered in terms of
the Poisson algebra.
$\!$Along the same as the Rajeev's method$\!$
\cite{Rajeev.94},
it is possible to construct the geometric quantization on 
the finite-dimensional Grassmannian
$\!\frac{SO(2N\!+\!2)}{U(N\!+\!1)}\!$.
According to Ref.
\cite{Nishi.Provi.08}, 
if we take matrix elements of ${\cal Q}$ and $\overline{{\cal Q}}$ 
as coordinates 
on the $\frac{SO(2N+2)}{U(N+1)}$ coset manifold,
the real line element is defined by a hermitian metric tensor 
on the coset manifold as
$
ds^2
\!\!=\!\!
\omega_{pq}{}_{\underline{r}~\!\!\underline{s}}
d{\cal Q}^{pq}d\overline{{\cal Q}}^{\underline{r}~\!\!\underline{s}}
({\cal Q}^{pq} \!=\! {\cal Q}_{pq},
\overline{{\cal Q}}^{\underline{r}~\!\!\underline{s}} 
\!\!=\!\!
\overline{{\cal Q}}_{\underline{r}~\!\!\underline{s}};~
\omega_{pq}{}_{\underline{r}~\!\!\underline{s}}
\!\!=\!\!
\omega_{\underline{r}~\!\!\underline{s}}{}_{pq}) .
$
We also use the indices
$\underline{r},~\underline{s},~\cdots$ 
running over 0 and $\alpha,~\beta,~\cdots$.
The hermitian metric tensor 
$\omega_{pq}{~}_{\underline{r}\underline{s}}$
is given through a K\"{a}hler potential, 
$
{\cal K}({\cal Q}^\dag,{\cal Q}) 
\!\!=\!\!
\ln \det \!
\left( \!
1_{\!N\!+\!1} 
\!+\! 
{\cal Q}^\dag \! {\cal Q} \!
\right) 
$
and its expression is given as 
$
\omega_{pq}{}_{\underline{r}~\!\!\underline{s}}
\!=\!
\frac{\partial ^2 {\cal K}({\cal Q}^\dag,{\cal Q})}
{\partial {\cal Q}^{pq} 
 \partial \overline{{\cal Q}}^{\underline{r}~\!\!\underline{s}}} .
$
Concerning the Poisson bracket, 
see textbook, e.g.,
\cite{HouHou.97}.
The Poisson bracket on the complex manifold for a pair of functions $f$ and $g$
is defined by
$\{f,g\} \!=\! -\omega^{-1}_{pq}{}_{\underline{r}~\!\!\underline{s}} \!\!
\left( \!
\frac{\partial  f}{\partial {\cal Q}^{pq}}
\!\otimes\!
\frac{\partial  g} 
{ \partial \overline{{\cal Q}}^{\underline{r}~\!\!\underline{s}}} 
\!-\!
\frac{\partial f} 
{ \partial \overline{{\cal Q}}^{\underline{r}~\!\!\underline{s}}}
\!\otimes\!
\frac{\partial  g}{\partial {\cal Q}^{pq}} \!
 \right) \!
 $
which is antisymmetric and satisfies the Jacobi identity.
The Hamiltonian operated on the function $\chi$,
the part of the $SO(2\!N\!+\!1)$ spinor function,
is the operator $\widetilde{\H}\!(\overline{\cal Q})$
coinciding with 
$
\frac{H(\overline{\cal Q},{\cal Q})}{S(\overline{\cal Q},{\cal Q})} .
$
While
we have the relation
$
H(\overline{\cal Q},{\cal Q})
\!\!=\!\!
\frac{H(G,G)}{|\Phi_{00}(G)|^2}
\!\equiv\!
<\!\!\! H \!\!\!>_{G,G} 
$
whose explicit expression,
using
(\ref{Hamiltonian}),
is obtained as
$
<\!\!\! H \!\!\!>_{G,G}
=\!\!
h_{\alpha\beta }{\cal R}_{\beta \alpha} 
\!\!+\!\!
\frac{1}{2}[\alpha\beta|\gamma\delta] \!
\left( \!
{\cal R}_{\beta \alpha}{\cal R}_{\delta \gamma}
\!-\! 
\frac{1}{2}
\overline{\cal K}_{\alpha\gamma }{\cal K}_{\delta\beta } \!
\right) 
$
and the SCF HB hamiltonian
$<\!\!\! H^{\mbox{\scriptsize{HB}}} \!\!\!>_{G,G}$ as
$
<\!\!\! H^{\mbox{\scriptsize{HB}}} \!\!\!>_{G,G}
=\!\!
{\cal F}_{\alpha\beta }{\cal R}_{\beta \alpha} 
\!+\!
\frac{1}{2}{\cal D}_{\alpha\beta }\overline{\cal K}_{\alpha\beta }
\!-\! 
\frac{1}{2}\overline{\cal D}_{\alpha\beta }{\cal K}_{\alpha\beta }
$
where the SCF parameters ${\cal F}_{\alpha\beta }$ and ${\cal D}_{\alpha\beta }$
are defined as
$
{\cal F}_{\alpha\beta }
\!\equiv\!
h_{\alpha\beta }
\!+\!
[\alpha\beta|\gamma\delta]{\cal R}_{\delta \gamma}
$
and
$
{\cal D}_{\alpha\beta }
\!\equiv\!
\frac{1}{2}[\alpha\gamma|\beta\delta]{\cal K}_{\delta\gamma} ,
$
respectively.
Both the
$<\!\!\! H \!\!\!>_{G,G}$
and
$<\!\!\! H^{\mbox{\scriptsize{HB}}} \!\!\!>_{G,G}$
become the quadratic functions of the $w$ through 
${\cal W}\!=\!\frac{1}{2}\left(w\!+\!1_{\!2N\!+\!2}\right)$.
These are the classical systems having the two-form $\omega$ under the constraint
$w^2 = 1_{\!2N\!+\!2}$.
In a near future, we will give a geometric requantization of the above classical systems,
following the ways, e.g., Hurt's
\cite{Hurt.83}
or
Kirillov's
\cite{Kirillov.88}.


\newpage

\vskip0.5cm

\noindent
\centerline{\bf Acknowledgements}

\vspace{0.5cm}

One of the authors (S.N.) would like to
express his sincere thanks to
Professor Constan\c{c}a Provid\^{e}ncia for
kind and warm hospitality extended to him at
the Centro de F\'\i sica, Universidade de Coimbra.
This work was supported by FCT (Portugal) under the project
CERN/FP/83505/2008.


\newpage


\end{document}